\documentclass[onecolumn,
               superscriptaddress,
               floatfix,
               longbibliography, 
               showkeys,apl]{revtex4-2}
\usepackage[utf8]{inputenc}
\usepackage{graphicx} 
\usepackage{bm}	
\usepackage{color}                     
\usepackage{xcolor}
\usepackage{epsfig}
\usepackage{amsmath} 
\usepackage{amssymb} 

\usepackage{float}
\usepackage{mathtools}
\usepackage{xparse}
\usepackage{hyperref}
\usepackage{chemformula}
\usepackage{physics}

\usepackage[frozencache,cachedir=.]{minted}
\usepackage{tcolorbox}
\tcbuselibrary{minted,breakable,xparse,skins}

\definecolor{bg}{gray}{0.95}
\DeclareTCBListing{mintedbox}{O{}m!O{}}{%
  breakable=true,
  listing engine=minted,
  listing only,
  minted language=#2,
  minted style=default,
  minted options={%
    linenos,
    gobble=0,
    breaklines=true,
    breakafter=,,
    fontsize=\small,
    numbersep=8pt,
    #1},
  boxsep=0pt,
  left skip=0pt,
  right skip=0pt,
  left=25pt,
  right=0pt,
  top=3pt,
  bottom=3pt,
  arc=5pt,
  leftrule=0pt,
  rightrule=0pt,
  bottomrule=2pt,
  toprule=2pt,
  colback=bg,
  colframe=orange!70,
  enhanced,
  overlay={%
    \begin{tcbclipinterior}
    \fill[orange!20!white] (frame.south west) rectangle ([xshift=20pt]frame.north west);
    \end{tcbclipinterior}},
  #3}

\usepackage[margin=1.25in]{geometry}

\usepackage[caption=false]{subfig}

\begin{document}

\title{ORQVIZ: Visualizing High-Dimensional Landscapes\\ in Variational Quantum Algorithms}

\author{Manuel S. Rudolph}
\affiliation{Zapata Computing Canada Inc., 325 Front St W, Toronto, ON, M5V 2Y1}

\author{Sukin Sim}
\affiliation{Zapata Computing, Inc., 100 Federal Street, Boston, MA 02110, USA}

\author{Asad Raza}
\affiliation{Zapata Computing Canada Inc., 325 Front St W, Toronto, ON, M5V 2Y1}

\author{Micha\l{} St\c{e}ch\l{}y}
\affiliation{Zapata Computing Canada Inc., 325 Front St W, Toronto, ON, M5V 2Y1}

\author{Jarrod~R.~McClean}
\affiliation{Google Quantum AI, 340 Main Street, Venice, CA 90291, USA}

\author{Eric~R.~Anschuetz}
\affiliation{MIT Center for Theoretical Physics, 77 Massachusetts Avenue, Cambridge, MA 02139, USA}

\author{Luis Serrano}
\affiliation{Zapata Computing Canada Inc., 325 Front St W, Toronto, ON, M5V 2Y1}

\author{Alejandro Perdomo-Ortiz}
\email{Correspondence: alejandro@zapatacomputing.com}
\affiliation{Zapata Computing Canada Inc., 325 Front St W, Toronto, ON, M5V 2Y1}

\date{\today}

\begin{abstract}
    Variational Quantum Algorithms (VQAs) are promising candidates for finding practical applications of near to mid-term quantum computers. 
    There has been an increasing effort to study the intricacies of VQAs, such as the presence or absence of barren plateaus and the design of good quantum circuit ansätze. Many of these studies can be linked to the loss landscape that is optimized as part of the algorithm, and there is high demand for quality software tools for flexibly studying these loss landscapes. In our work, we collect a variety of techniques that have been used to visualize the training of deep artificial neural networks and apply them to visualize the high-dimensional loss landscapes of VQAs. We review and apply the techniques to three types of VQAs: the Quantum Approximate Optimization Algorithm, the Quantum Circuit Born Machine, and the Variational Quantum Eigensolver. Additionally, we investigate the impact of noise due to finite sampling in the estimation of loss functions. 
    For each case, we demonstrate how our visualization techniques can verify observations from past studies and provide new insights. 
    This work is accompanied by the release of the open-source Python package \textit{orqviz}, which provides code to compute and flexibly plot 1D and 2D scans, Principal Component Analysis scans, Hessians, and the Nudged Elastic Band algorithm. 
    \textit{orqviz} enables flexible visual analysis of high-dimensional VQA landscapes and can be found at: \href{https://github.com/zapatacomputing/orqviz}{\textbf{github.com/zapatacomputing/orqviz}}.
\end{abstract}

\maketitle

\section{Introduction}
Variational Quantum Algorithms (VQAs) are promising candidates for finding practical applications of near to mid-term quantum computers \cite{cerezo2020variational, bharti2021noisy}. They can be used, for example, to solve optimization problems, implement machine learning models, find the ground state of quantum systems, and more. Despite the range of applications, it remains difficult to derive performance guarantees for VQAs.
In classical machine learning, in particular deep learning, model architecture and optimization (training) procedures are treated as black boxes, many of which lack theoretical guarantees, or their performance guarantees are uninformative in practice. For instance, optimizing deep neural networks is in general NP-hard~\cite{BLUM199NPhard}. However, naive gradient descent optimizers are able to effectively train modern neural networks, and additionally, the optimization trajectories are often very low-dimensional~\cite{li2017visualizing}. Current VQAs do not parallel classical deep learning algorithms in terms of the number of parameters or complexity of the ansatz, and now is the best time to build strong foundational understanding of available and upcoming approaches. Practitioners can evaluate and record many quantitative metrics to generate practical heuristics, however, visualization is a natural tool for building intuition and revealing promising directions of research. It can improve understanding of the optimization task, the method used for optimization, or the design quantum circuits. We definite \textit{visualization} as the practice of depicting or probing the shape of the high-dimensional loss landscape in low-dimensional plots.\\\\

The aim of this work is to review several techniques to visualize landscapes of VQAs. We apply these techniques on a range of examples to better understand their capabilities and limitations. This work is accompanied by the release of \textit{orqviz}, an open-source Python package for visualizing VQA landscapes that supports the techniques described in this work and more. 
The example cases are in the domains of combinatorial optimization, generative machine learning, and molecular ground state search. Additionally, we study the effect of shot noise on the loss landscapes and optimization of parameterized quantum circuits. 
Our general ethos is to perform \textit{conventional} quantitative studies and then extend our understanding with visualization tools. In the following, we briefly describe the VQA examples studied in this work.

The \textit{Quantum Approximate Optimization Algorithm} (QAOA)~\cite{farhi2014quantum} is a quantum heuristic used for solving combinatorial optimization problems. The optimization task is equivalent to finding the ground state of a diagonal Ising Hamiltonian, which encodes the combinatorial problem. Due to the quantum resources required by QAOA, as well as challenges related to vanishing gradients, also called \textit{barren plateaus}~\cite{mcclean2018barren}, good parameter initialization for VQAs including QAOA is crucial.
A phenomenon called \textit{parameter concentration}~\cite{brandao2018concentration} promises to provide high-quality initial guesses by solving a similar but much smaller problem instance and transferring those parameters to the problem of interest. We study and confirm parameter concentration by comparing the loss landscapes of unweighted and weighted 3-regular graphs of up to 20 qubits. Additionally, we visually confirm parameter concentration in different instances of the graph-partitioning problem (see Appendix~\ref{apx:QAOA_more}), indicating that parameter concentration is a widespread phenomenon in QAOA.

The \textit{Quantum Circuit Born Machine} (QCBM)~\cite{Benedetti2019generative} is a generative machine learning algorithm that encodes a probability distribution over discrete data in the measurement probabilities of a quantum wavefunction. Generative modeling tackles the task of learning the probability distribution that underlies available data in order to sample from the distribution and generate similar but new data. This concept is called \textit{generalization} and it is vitally important that we  understand the generalization capabilities of quantum generative models. Here, we study the loss landscapes of the QCBM trained with different amount of training data to understand how the structure of the loss landscape is linked to the generalization performance of the model.

The \textit{Variational Quantum Eigensolver} (VQE)~\cite{peruzzo2014variational} is a near-term algorithm for approximating the ground state energy of a Hamiltonian. 
It is well-known that VQE instances can be very hard to optimize, often requiring highly specific ansätze and parameter initialization
\cite{romero2019strategies, mcclean2018barren, wecker2015progress, dallairedemers2019low, lee2019generalized, anschuetz2021critical}. In this work, we compare two particular VQE ansätze, UCCSD and $k$-UpCCGSD, for estimating the ground state energy of the $H_4$ molecular system using visualization techniques. 
We confirm that $k$-UpCCGSD is the more expressive ansatz even at low $k$ values, and that the ansatz parameters can be \textit{continuously} manipulated to estimate the ground state energies for various bond lengths, unlike for UCCSD.
However, the increase in flexibility in $k$-UpCCGSD comes at the cost of a greater number of minima that are separated by relatively high energy barriers, leading to sensitivity in parameter initialization.

Sections of this work are organized as follows:\\
\begin{itemize}
    \item Section~\ref{sec:example} motivates the importance of visualization in understanding the loss landscape of a parameterized algorithm by considering a toy loss function. Here, we present and apply a variety of visualization techniques (using \textit{orqviz}) to analyze the toy function and demonstrate how results from these techniques can inform the nature and characteristics of the toy function.
    For more details on the techniques, we refer the readers to Appendix~\ref{apx:viz_techniques}.
    \item Section~\ref{sec:QAOA} demonstrates how visualization techniques can be applied to study and extend our understanding of the \textit{parameter concentration} phenomenon in QAOA.
    \item Section~\ref{sec:QCBM} studies the notions of \textit{generalization} and training with limited data in the context of generative machine learning with QCBMs.
    \item Section~\ref{sec:VQE} presents a visual comparison of two particular ansätze, UCCSD and $k$-UpCCGSD, for VQE applied to the $H_4$ molecular system. 
    \item Section~\ref{sec:NOISE} presents the impact of finite-sampling noise on the training of VQAs. Namely, we observe and compare how noise from finite-sampling could impact the loss landscapes and training of VQE and QCBM.
\end{itemize}

\section{Visualization Techniques - A brief overview}\label{sec:example}
To introduce the reader to several of the visualization techniques reviewed in this work and supported by \textit{orqviz}, we consider the following toy loss function as a case study:
\begin{equation}\label{eq:sombrero}
    \mathcal{L}(\bm\theta) = 1 - \frac{\sin(\nu \|\bm\theta\|)}{\nu \|\bm\theta\|}
\end{equation} 
with parameter $\nu=2$ and $\|\cdot\|$ being the $2$-norm. Even though the loss function is analytical and known \textit{a priori}, we demonstrate how results from various visualization techniques come together to create a richer picture of its loss landscape. \\\\

Given the loss function $\mathcal{L}(\bm\theta)$ that is parameterized by a high-dimensional parameter vector $\bm\theta$, how can one try to learn more about the loss landscape of $\mathcal{L}$?
One common practice is to optimize the parameterized model repeatedly, starting from various random initial points and evaluating the loss as a function of the training iterations. Through such numerical simulations, we additionally obtain information about the variance and quality of final minima. 
In this work, we refer to this method as a type of ``quantitative analysis''.
While these quantities can be valuable in probing the landscape, the information we gain about the problem is still limited.
Another approach to studying the loss landscape of interest can be theoretically, where one derives properties from the definition of the algorithm. It may however not be well known how these properties manifest in practice. Given that we know the loss function, we could make predictions about aspects of the optimization and test them. However, in this section, we approach the task from a visualization perspective.

For this example case, the loss function $\mathcal{L}(\bm\theta)$ is parameterized by a $4$-dimensional parameter vector $\bm\theta$. Consequently, the loss function does not have a single representation in 2D or 3D that could fully explain the loss landscape. As an initial numerical study, we present 100 repetitions of 150 iterations with gradient descent, starting optimization from random parameters, each sampled uniformly in the interval $[-5, 5]$. 
\begin{figure}[H]
\centering
\includegraphics[width=0.65\linewidth]{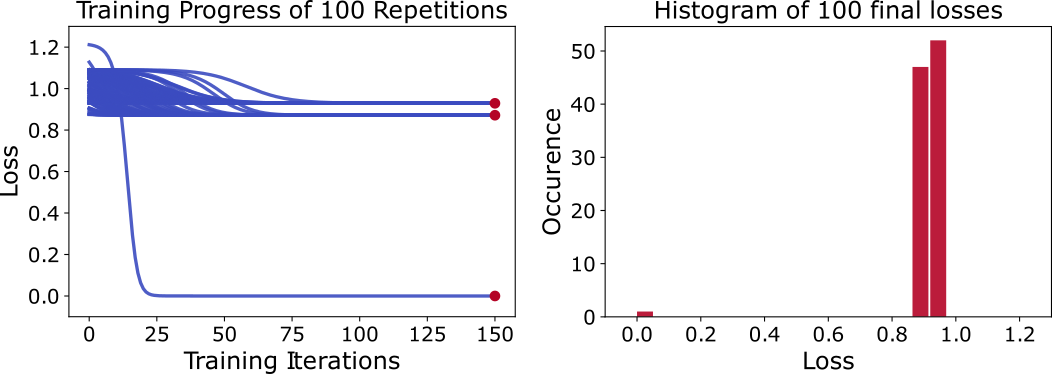}
\end{figure}
After around 75 iterations of gradient descent, all instances converge close to their final losses. This quantitative analysis allows resolving the distribution of local minima, which appears to be very discrete. Only one out of 100 repetitions achieved the (likely) global minimum with $\mathcal{L}_0(\bm\theta)=0$. All other runs converge to one of two final losses, $\mathcal{L}_1=0.872$ and $\mathcal{L}_2=0.929$. This tells us that the loss landscape is mostly regions of local minima and has a narrow region of global minimum. While this is valuable information,
there is still much to uncover about the loss landscape. The loss progression over the training iterations does not generally reflect the shape of the loss landscape, especially not for non-local optimizers.

To gain a deeper understanding of the nature and characteristics of loss landscapes, we now present
different visualization techniques and what each reveals about the example loss function. More detailed descriptions of the visualization techniques can be viewed in Appendix~\ref{apx:viz_techniques}.

Many of the visualization techniques presented in this work involve \textit{scanning} the loss landscape. These scans can be 1D or 2D, i.e. along a path or a surface, respectively. A scan $f$ is not a projection, rather a \textit{slice} of the loss landscape, where every parameter vector $\bm\theta^\prime$ on a 1D or 2D manifold $\mathcal{M}$ is evaluated via $f(\mathcal{M}) = \{\mathcal{L}(\bm\theta^\prime)\}_{\bm\theta^\prime \in \mathcal{M}}$. In this case, visualization then revolves around how to choose the manifold $\mathcal{M}$ and how to present or plot the function $f(\mathcal{M})$. In some cases, visualization techniques give direct, potentially quantitative information about the loss landscape. We note that the ability to visualize high-dimensional landscapes of VQAs is not limited to the loss function. With \textit{orqviz}, any function that takes parameter vectors as input and returns a real number can be studied. For instance, one could scan fidelity with the (known) ground state of a Hamiltonian, variance of the loss or gradient estimate, test error in a machine learning task, and more. In this section, we will restrict ourselves to studying the loss landscape.

\subsubsection*{Interpolation Scans}
In the case of \textit{interpolation scans}, the loss function is evaluated along the direction vector $\textbf{d} = \bm\theta_2 - \bm\theta_1$. Interpolation scans can, for example, be applied between two random points or previously found solutions of optimization procedures. They can be performed in 1D or 2D, where the second direction vector can be chosen arbitrarily, for example orthonormal to \textbf{d}. In the following, we perform interpolations scans between minima on the three different levels of local minima. More details on the fundamental scans such as interpolation scans can be viewed in Appendix~\ref{apx:simple_scans}.
\begin{figure}[H]
\centering
\includegraphics[width=0.7\linewidth]{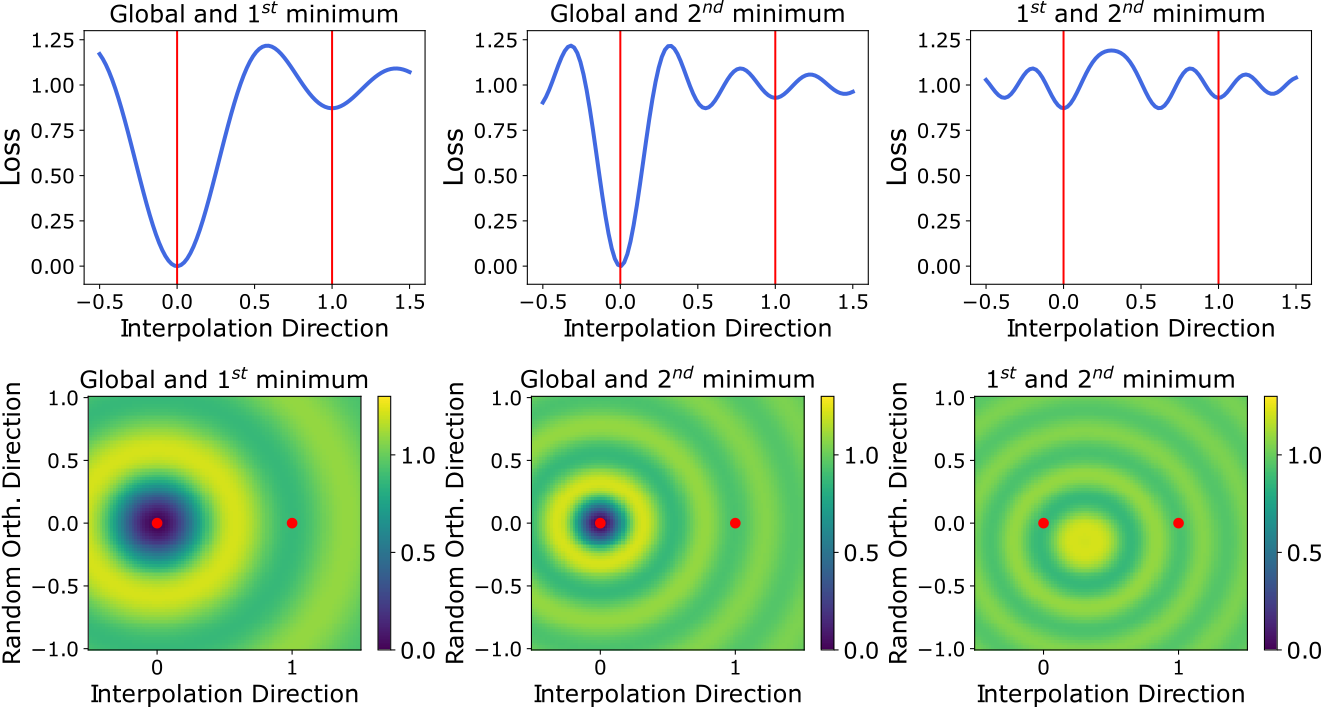}
\end{figure}
The 1D scans reveal oscillations in the loss landscape which were not apparent from the quantitative analysis. Such 1D scans are not very expensive to simulate, they can however miss relevant features. Performing 2D scans instead uncovers that the oscillations are in fact concentric rings. Even though the second direction was chosen to be random and orthogonal to $\textbf{d}$, we are able to clearly resolve this property of the landscape. \textit{orqviz} supports highly flexible customization of the scans performed. A basic version of these results can be generated using:
\begin{samepage}
\begin{mintedbox}{python}
from orqviz.scans import perform_1D_interpolation, plot_1D_interpolation_result

interpolation1D_result = perform_1D_interpolation(minimum1_parameters, minimum2_parameters, loss_function=loss_function)
plot_1D_interpolation_result(interpolation1D_result)

from orqviz.scans import perform_2D_interpolation, plot_2D_interpolation_result

interpolation2D_result = perform_2D_interpolation(minimum1_parameters, minimum2_parameters, loss_function=loss_function)
plot_2D_interpolation_result(interpolation2D_result)
\end{mintedbox}
\end{samepage}

\subsubsection*{Principal Component Analysis}
\textit{Principal Component Analysis} (PCA)~\cite{Pearson1901pca} is a deterministic linear algorithm for dimensionality reduction. In our work, it is especially relevant for choosing spanning vectors for scans. When applying the PCA algorithm to a collection of high-dimensional vectors, it outputs a hierarchy of orthogonal directions in which the collection of parameter vectors display the largest variance. These directions are called \textit{components}. By choosing the first two components as spanning vectors for the manifold $\mathcal{M}$, one performs a 2D scan which is the most representative of the original collection of parameter vectors. The parameter vectors can then be projected onto the 2D surface with the smallest approximation error. Consequently, the utility of PCA scans is limited by the dimensionality of the point collection. If the first two components do not explain a significant fraction of total variance in a high-dimensional space (for example upwards of $>60\%$), the scanned loss landscape is not particularly representative of the points. However, throughout our studies, we have observed surprisingly low-dimensional optimization trajectories in VQA problems, with upwards of $80\%$ explained by the first two components, making PCA scans a very useful tool. They are used in every section throughout this work. 
\begin{figure}[H]
\centering
\includegraphics[width=0.7\linewidth]{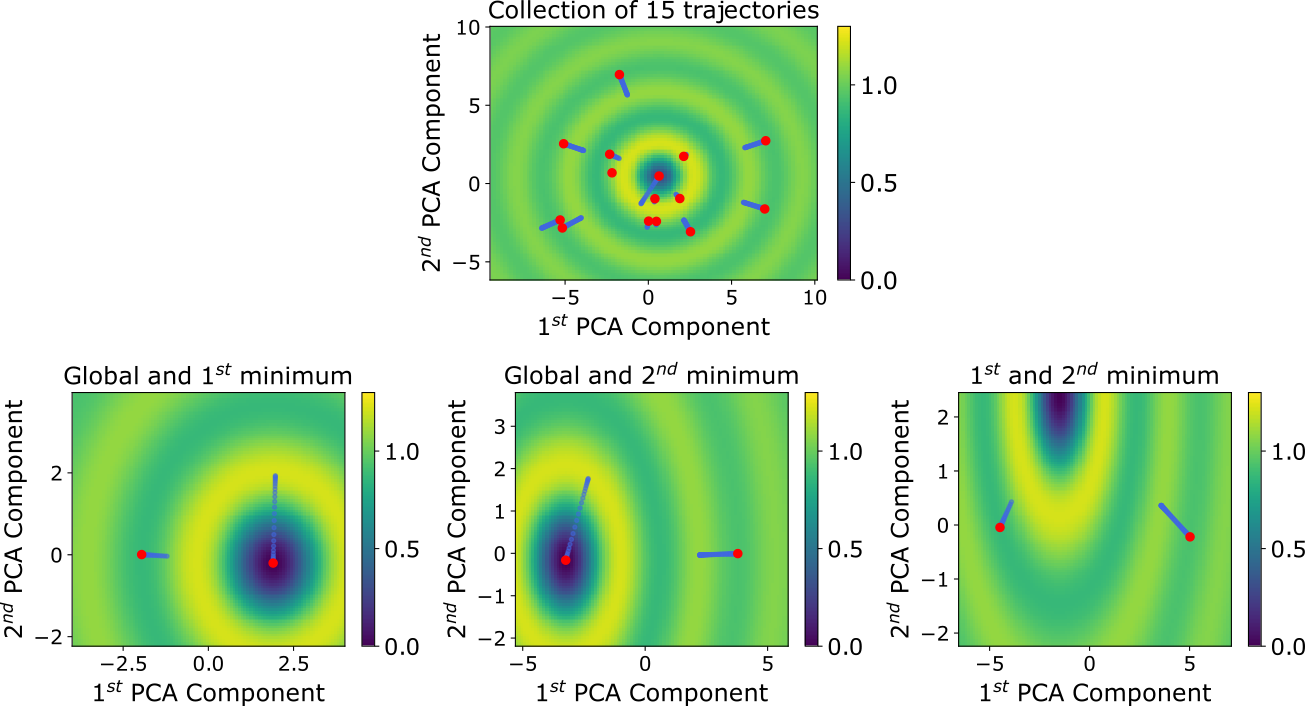}
\end{figure}
In the case of our toy example, the parameter trajectories arising from optimization with gradient descent are exactly linear. This indicates that the loss landscape does not have intricate structure that the scans are missing and optimization runs simply converge to the closest concentric ring. Using \textit{orqviz}, PCA scans and trajectory projection can be performed via:
\begin{samepage}
\begin{mintedbox}{python}
from orqviz.pca import get_pca, perform_2D_pca_scan, plot_pca_landscape, plot_optimization_trajectory_on_pca

pca_object = get_pca(parameter_trajectory)
pca_scan_result = perform_2D_pca_scan(pca_object, loss_function)
plot_pca_landscape(pca_scan_result, pca_object)
plot_optimization_trajectory_on_pca(parameter_trajectory, pca_object)
\end{mintedbox}
\end{samepage}

\subsubsection*{Hessian}
The \textit{Hessian} of a loss function is the matrix of second partial derivatives of that loss function with respect to the model parameters. As such, the Hessian indicates local curvature, i.e. how quickly the gradient of the loss changes with respect to changes in parameters. The Hessian can be expensive to estimate, as exact calculation scales quadratically with the number of parameters. However, the eigenvalues and eigenvectors of the Hessian are highly informative. In cases where scans of the loss landscape alone do not explain certain phenomena, the Hessian can be used to extract additional information. The eigenvalues can either be used to inspect the widths of minima, or the respective eigenvectors as spanning directions for 1D or 2D scans. Consequently, one can perform scans with direction vectors in which the loss increases particularly slowly or quickly. Additionally, as mentioned above, one is not limited to scanning loss function values. The ratio $\frac{\lambda_{min}}{\lambda_{max}}$ of the smallest ($\lambda_{min}$) to largest ($\lambda_{max}$) eigenvalue of the Hessian can be scanned instead to reveal the curvature of the loss landscape. This technique is very resource intensive; however, we utilize it in Sec.~\ref{sec:QCBM} to resolve detailed differences between two landscapes. More details on visualization with the Hessian can be viewed in Appendix~\ref{apx:hessian}.
\begin{figure}[H]
\centering
\includegraphics[width=0.7\linewidth]{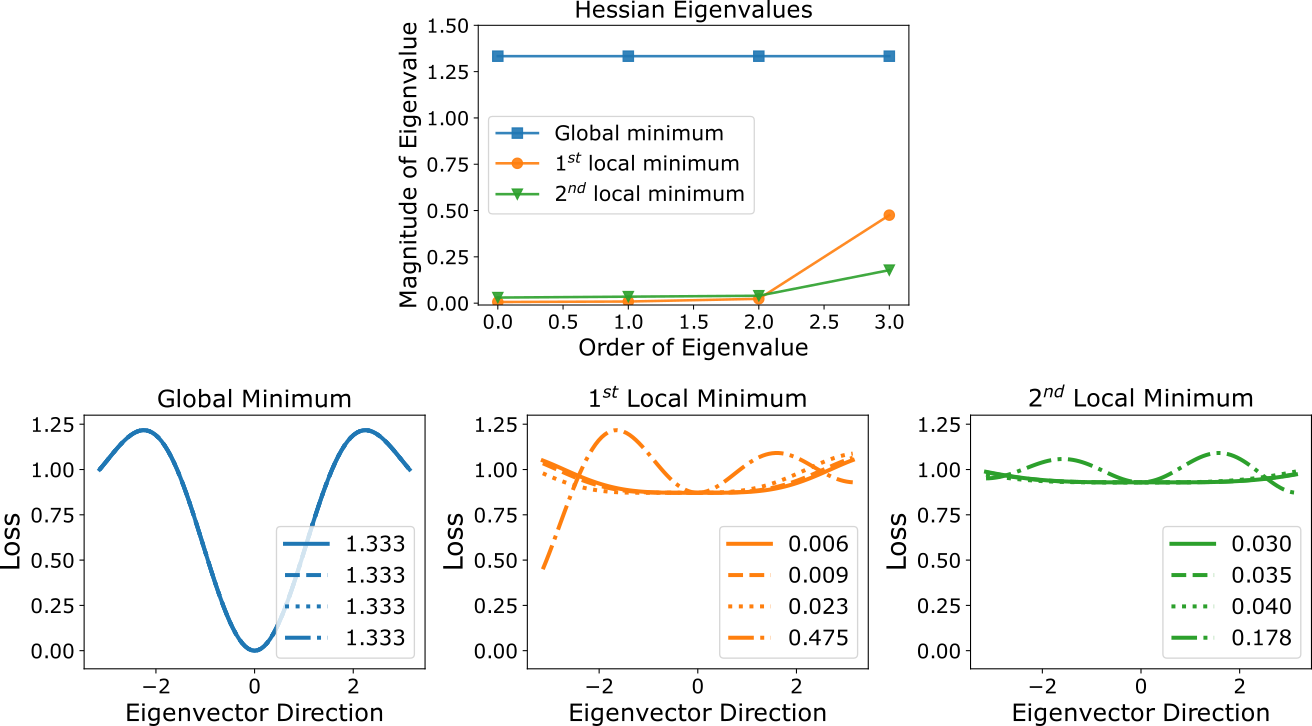}
\end{figure}
Here, we show the distribution of eigenvalues of the Hessians for different minima which are all non-negative, indicating that the landscape is locally convex. The eigenvalues also reveal that the global minimum lies at the bottom of a steep basin, while the local minima have at least one eigenvalue very close to zero. This implies that the corresponding direction is locally ``flat''. When performing 1D scans along the eigenvector directions (plot labels indicate the corresponding eigenvalues), we see that the loss, while flat locally, does increase slowly. These observations are consistent with the previous visualizations of the concentric rings. \textit{orqviz} supports a full numeric calculation of the Hessian, as well as an approximation that is based on stochastic gradient approximation (see Appendix~\ref{apx:hessian} for details). For applications in which utilizing the Hessian can be useful, scanning in the direction of Hessian eigenvectors can be performed like this:
\begin{samepage}
\begin{mintedbox}{python}
from orqviz.hessians import get_Hessian, perform_1D_hessian_eigenvector_scan, plot_1D_hessian_eigenvector_scan_result

hessian = get_Hessian(parameter_vector, loss_function)
list_of_1Dscans = perform_1D_hessian_eigenvector_scan(hessian, loss_function)
plot_1D_hessian_eigenvector_scan_result(list_of_1Dscans)
\end{mintedbox}
\end{samepage}

\subsubsection*{Nudged Elastic Band}
The \textit{Nudged Elastic Band} (NEB)~\cite{jonsson1998nudged} is an algorithm which can be used to find low-loss paths in the landscapes of VQAs. It has been used to reveal that pairs of minima in deep  neural networks are often connected by paths of very low loss~\cite{draxler2019NEB}. This finding has been essential to our understanding about the trainability of artificial neural networks. Historically, NEB has also been widely used in chemistry to find transition states and minimum energy paths for reactions~\cite{Jonsson2000improved}. The NEB represents a piece-wise linear path through the high-dimensional space and consists of a collection of pivot points connected by springs with a specific spring constant. The pivots follow the local gradient of the loss function, as well as the restoring force of the springs. Intuitively, the elastic band slopes in the loss landscape just like a rope, being held horizontally by the ends, slopes under the gravitational force. More details on the NEB algorithm can be viewed in Appendix~\ref{apx:elastic_band}.
\begin{figure}[H]
\centering
\includegraphics[width=0.55\linewidth]{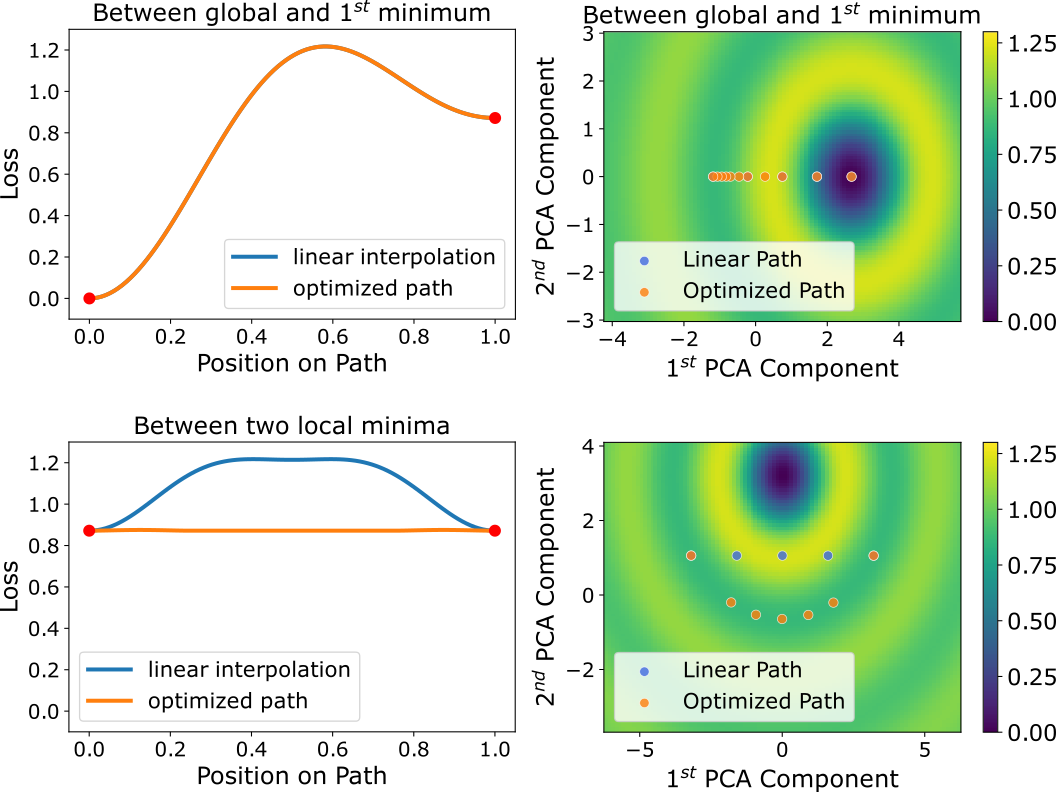}
\end{figure}
We apply the NEB between local minima to study how the different minima are connected. This reveals that there is a significant loss barrier to reach the global minimum which cannot be avoided. However, two different minima  (based on the Euclidean distance between the parameters) at the same loss are connected by a low-loss path. Arguably, every solution in the same ring of the landscape lies in the same minimum. PCA scans and the NEB algorithm can naturally be combined to reveal how the path curves around the global minimum.

Ref.~\cite{Kolsbjerg2016AutoNEB} presents an automatic extension of the NEB algorithm, called AutoNEB. The algorithm performs gradient descent optimization on the path in cycles, checking for and adding new pivots where necessary in-between the cycles~(see Appendix~\ref{apx:elastic_band} for details). \textit{orqviz} implements both algorithms, NEB and AutoNEB. For example, the advanced AutoNEB algorithm can simply be implemented for any VQA application via:
\begin{samepage}

\begin{mintedbox}{python}
from orqviz.NEB import Chain, ChainPath, run_AutoNEB

linear_chain = Chain(np.linspace(minimum1, minimum2, num=5))
all_chains = run_AutoNEB(linear_chain, loss_function, n_cycles=n_cycles, n_iters_per_cycle=n_iters_per_cycle, max_new_pivots=max_new_pivots, learning_rate=learning_rate)
trained_chain = all_chains[-1] 
\end{mintedbox}
\end{samepage}

\begin{figure}[H]
\centering
\includegraphics[width=0.40\linewidth]{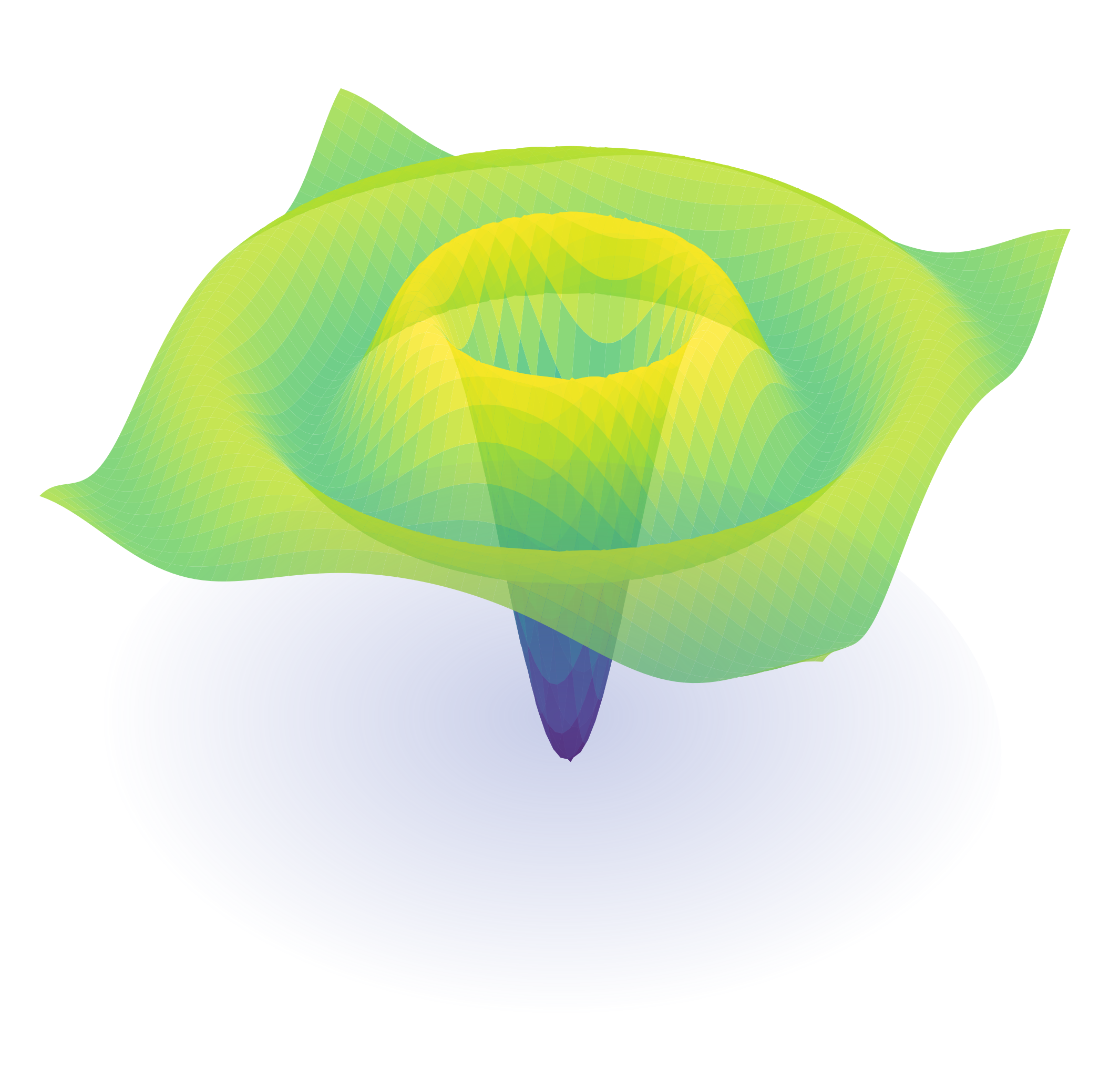}
\caption{A 2D scan of the ``Sombrero'' function that was plotted using the ``\textit{plot\_2D\_scan\_result\_as\_3D}'' method from \textit{orqviz}.}
\label{fig:sombrero}
\end{figure}

All of the visualization techniques we reviewed can be seamlessly combined with each other to produce a richer picture of the landscape. From applying our techniques, we know:
\begin{enumerate}
    \item The landscape has different levels of losses, with one significantly lower loss at 0. From our numerical simulations, we also know that this low loss is in a narrower region, as the probability of reaching that minimum is low. 
    \item The interpolation and PCA scans inform us that the loss landscape has concentric rings. At the center of the rings is the minimum with the lowest loss.
    \item The Hessian matrix at all minima is positive semi-definite (all eigenvalues are non-negative), confirming that the landscape is locally convex. Additionally, the higher-loss minima show a locally ``flat'' direction.
    \item The NEB results show that minima of the same quality are connected by paths of low loss. This indicates that solutions with the same loss sit in the same concentric ring. 
\end{enumerate}
A brief discussion of the computational resources required for our visualizations can be viewed in Appendix~\ref{apx:comp_resources}. While the initial quantitative studies generated the necessary optimization trajectories and solution candidates that the visualization techniques were applied to, only the latter revealed that $\mathcal{L}$ is a radial function where the initial parameters fully determine the outcome of a gradient descent optimizer, and that local minima at the same loss are fully connected. In fact, Eq.~\ref{eq:sombrero} can now be understood as a variant of the ``Sombrero'' function. A 3D demonstration of the function can be seen in Fig.~\ref{fig:sombrero}.

One important consideration for studying parameterized loss landscapes are the model symmetries. In our example, the loss function only depends on the norm of the parameter vector $\bm\theta$. In fact, we can define a new dependent variable $r=||\bm\theta||$ to denote the radial distance to the origin. This reduces the problem to a 1-dimensional optimization task, removing the rings of degenerate solutions in the process. We emphasize the importance of careful inspection of the symmetries in the loss landscape, which has been shown to be critically important in the study of classical neural networks~\cite{li2017visualizing}. VQAs may contain a range of symmetries and one of the most general ones is the periodicity of the parameters that control operations in the quantum circuit. Depending on the specific convention, the outcome of parameterized quantum circuits is unaffected when adding integer multiples of $2\pi$ (or $4\pi$) to any or all parameters. Additionally, the path between parameter vectors in high-dimensional space depends heavily on whether it traverses along the real number line, or whether it utilizes the periodicity and \textit{wraps} around the period. For example, the shortest path between $2\pi$-periodic parameters $\theta_1 = 0.1$ and $\theta_2 = 2\pi - 0.1$ is along the vector $\text{d} = \theta_1 - \theta^\prime_2$ with $|\text{d}|=0.2$ where $\theta^\prime_2 = \theta_2 - 2\pi$. Neglecting these symmetries can cause unphysical artifacts in the study of the respective loss landscape. \textit{orqviz} provides helper functions to wrap parameter vectors or entire parameter trajectories according to their nearest periodic ``copy'' relative to a reference point. See Appendix~\ref{apx:periodic_wrap} for details and an \textit{orqviz} code example.

In the following sections, we apply a range of the visualization techniques presented in this section to examples of VQAs.

\section{QAOA - Visualizing parameter concentration}\label{sec:QAOA}

As mentioned before,  QAOA is a VQA for solving the ground state problem for a given diagonal Hamiltonian $H_C$, the \emph{cost Hamiltonian}, typically with the goal of solving a combinatorial optimization problem~\cite{farhi2014quantum}. Prior to optimization, a so-called \emph{mixer Hamiltonian} $H_B$ is chosen that does not commute with $H_C$, and has a ground state $\ket{\psi_0}$ which is easy to prepare. Typically, one uses $\ket{\psi_0}=\ket{+}^{\otimes n}$ and
\begin{equation}
    H_B=-\sum\limits_{i=1}^n\sigma_i^x.
\end{equation}
Here, $\sigma^\alpha$ ($\alpha=x,y,z$) correspond to the single-qubit Pauli matrices for an $n$-qubit state.
Then, given circuit depth $p$, an ansatz state
\begin{equation}
    \ket{\bm{\gamma},\bm{\beta}} = \prod\limits_{k=1}^p e^{-i\beta_k H_B} e^{-i\gamma_k H_C}\ket{\psi_0}
    \label{eq:qaoa_ansatz}
\end{equation}
is prepared, with parameters $\left(\bm{\gamma},\bm{\beta}\right)$ chosen to minimize the cost function
\begin{equation}
    \mathcal{L}\left(\bm{\gamma},\bm{\beta}\right)=\bra{\bm{\gamma},\bm{\beta}}H_C\ket{\bm{\gamma},\bm{\beta}}.
\end{equation}
The ansatz state given in Eq.~\eqref{eq:qaoa_ansatz} is motivated by adiabatic quantum computing~\cite{Nishimori1998annealing,Farhi2000annealing}.
\begin{figure}[t]
    \centering
    \includegraphics[width=0.95\linewidth]{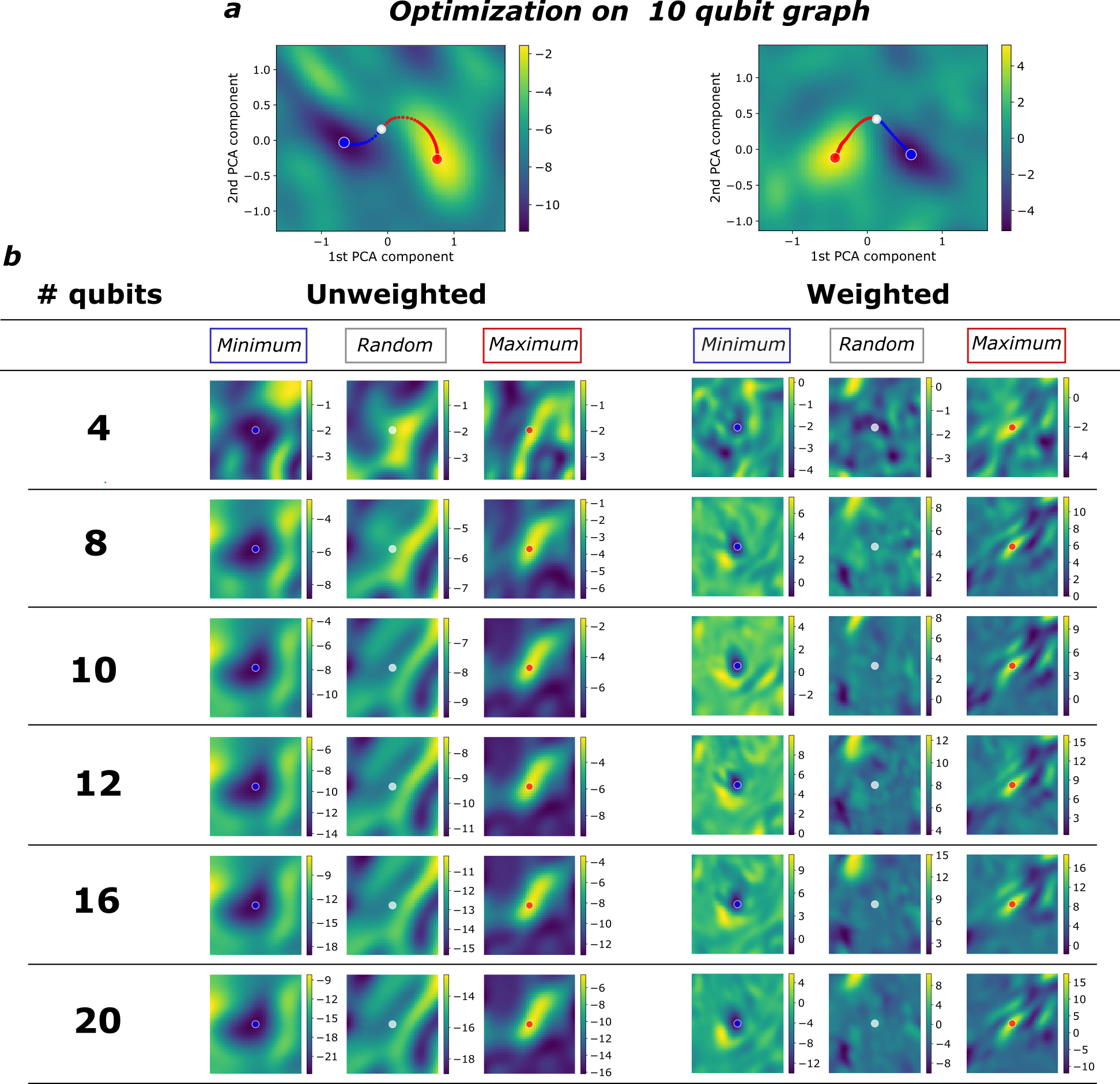}
    \caption{Visualization of parameter concentration in QAOA for \textsc{Max-Cut} on unweighed ($w_{ij}=1$) and weighted ($w_{ij}\in\left\{\pm 1,\pm 2,\pm 3\right\}$) 3-regular graphs. Layer depth for the QAOA ansatz is $p=4$ with eight parameters. \textbf{(a)} PCA scan of the optimization trajectories to find a local \emph{minimum} (blue) and local \emph{maximum} (red) points starting from \textit{random} initialization (white) on a single random 10 qubit graph instance. \textbf{(b)} 2D scans of the loss landscapes around the points marked with \emph{minimum},  \textit{random}, and \emph{maximum}. These points, as well as the same scan direction vectors, are used over all graph instances to get the landscape cross sections. Note the concentration of the loss landscapes, particularly at large numbers of qubits.}
    \label{fig:QAOA_scans}
\end{figure}

In this work, we consider the \textsc{Max-Cut} problem. In \textsc{Max-Cut}, the goal is to partition a graph into two subgraphs, such that the sum of weights over edges between the two subgraphs is maximized. This is equivalent~\cite{Lucas2014Ising} to finding the ground state of the cost Hamiltonian
\begin{equation}\label{eq:maxcut_hamiltonian}
    H_C = \sum\limits_{\left\langle i,j\right\rangle\in E}w_{ij}\sigma_i^z \sigma_j^z,
\end{equation}
where $w_{ij}$ is the weight of the edge $\left\langle i,j\right\rangle$. Here, we consider the \textsc{Max-Cut} problem on 3-regular graphs, where every vertex has nonzero edge weight on only three connecting edges. We consider both unweighted instances, i.e. all nonzero $w_{ij}=1$, and also weighted graph instances with random nonzero integer weights $w_{ij}\in\left\{\pm 1,\pm 2,\pm 3\right\}$.

One interesting phenomenon found to be present in the loss landscape of QAOA applied to 3-regular graphs is that of \emph{parameter concentration}~\cite{brandao2018concentration}. Namely, it has been observed that for given fixed parameters $\bm{\gamma},\bm{\beta}$, the loss $\mathcal{L}$ concentrates across problem instances. This can be applied to the optimization of the QAOA loss function for large instances, i.e. optimal parameters found at smaller problem instances can be used as the initial parameters in the optimization of QAOA applied to large problem instances. 
\begin{figure}[t]
    \centering
    \includegraphics[width=0.75\linewidth]{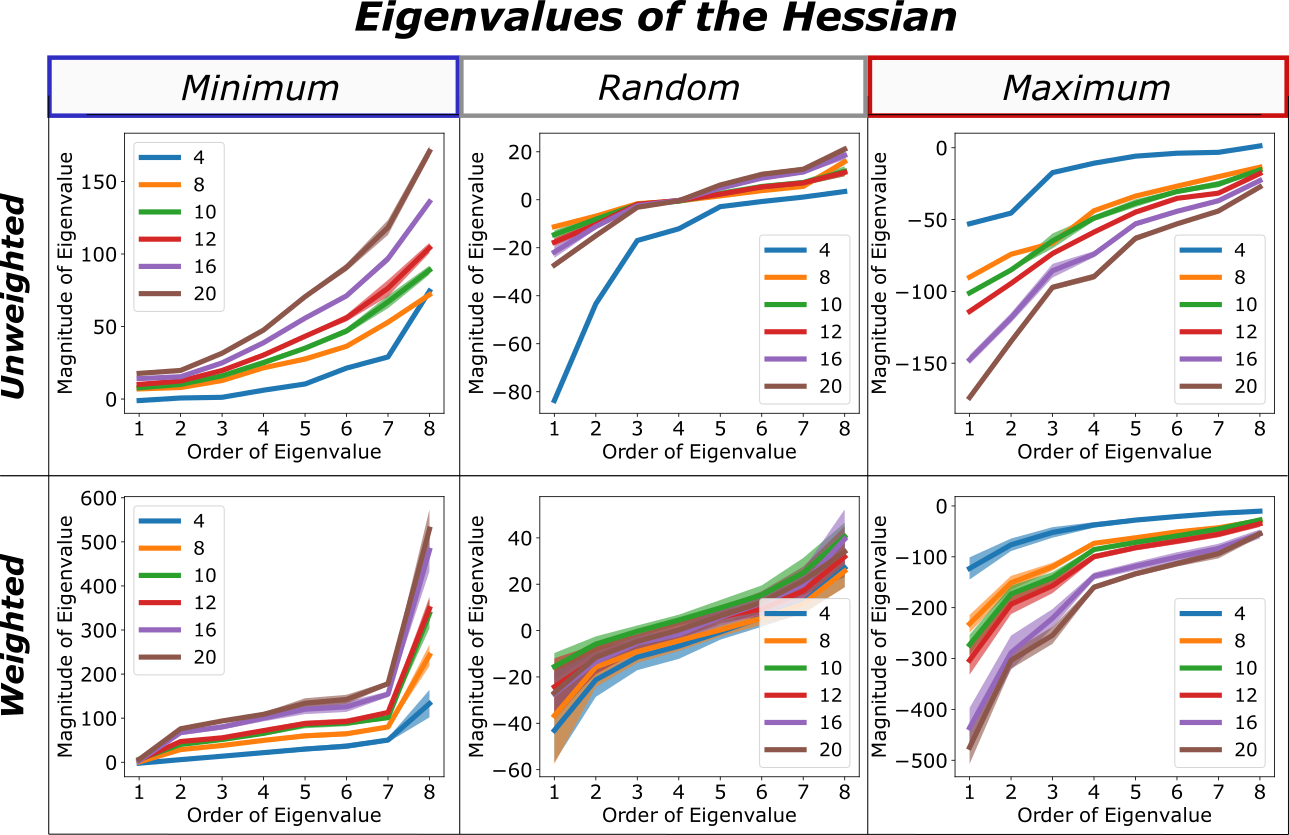}
    \caption{Hessian eigenvalues for a family of unweighed ($w_{ij}=1$) and weighted ($w_{ij}\in\left\{\pm 1,\pm 2,\pm 3\right\}$) 3-regular graph instances with different number of qubits. For all instances, the Hessian is estimated at the (local) \emph{minimum}, a \emph{random} point, and the (local) \emph{maximum} from Fig.~\ref{fig:QAOA_scans}. For these plots, the Hessian eigenvalues are collected over 10 random graph instances of each graph size. The solid lines represent the averages while the shaded regions mark one standard deviation. The variation of the curvature in the loss landscapes appears to be low between graph instances of the same size and weight range, particularly at the extrema.}
    \label{fig:QAOA_hessians}
\end{figure}
In the context of our work, we study the concentration of the loss landscape as a whole over various graph instances and problem sizes. Similarly to~\cite{brandao2018concentration}, we identify parameters $\bm{\gamma},\bm{\beta}$ that correspond to a local \textit{minimum}, a \textit{random} point, and a local \textit{maximum} for single instances of randomly chosen 3-regular unweighted and weighted 10 \textsc{Max-Cut} qubit graphs. The local minimum and local maximum were found by minimizing and maximizing the cost function, respectively, starting from a random set of parameters for $\bm{\gamma},\bm{\beta}$. To clarify how these points are found, see Fig.~\ref{fig:QAOA_scans}a for PCA scans on the optimization trajectories. The landscapes near the same points in parameter space are then observed on a variety of $4$ to $20$ qubit graphs. To test the similarity of the loss landscapes over these problem instances, we fix the spanning vectors $\textbf{v}_1, \textbf{v}_2$ for the 2D scans, which were randomly sampled from a normal distribution and have norm $\left\lVert\bm{v}_1\right\rVert=\left\lVert\bm{v}_2\right\rVert=1.5$.
Fig.~\ref{fig:QAOA_scans}b shows the results of the 2D scans near the parameters of interest which clearly demonstrate the phenomenon of parameter concentration over a variety of system sizes. For both the unweighted and weighted 3-regular graphs, the loss landscapes are nearly indistinguishable (up to an overall energy scaling) when varying the size of the graph instance. Similarly, we find that the loss landscape of graph instances of the same size concentrate as well. This can be seen in Fig.~\ref{fig:QAOA_hessians}, where we depict the mean and standard deviation of the Hessian eigenvalues at the same local extrema over 10 random graph instances per graph size. We observe low relative variance in curvature over different problem instances of the same size, particularly at the extrema. While the absolute losses differ significantly for the weighted graphs, the Hessian eigenvalues reveal that the curvature is similar across graphs. Additionally, we find a clear trend in the Hessian eigenvalues when scaling to large numbers of qubits which is due to the increasing magnitude of the maximal possible cut in larger graphs.

Our visualization results confirm the quantitative studies of~\cite{brandao2018concentration}, as well as more recent work utilizing this phenomenon~\cite{larkin2020evaluation, streif2020QAOA_noaccess, Zhou2020QAOAperformance, sauvage2021flip}. In fact, the degree of parameter concentration could be quantified using similarity metrics between visualization scans. The approach shown in Fig.~\ref{fig:QAOA_scans} can flexibly be applied to different types of graphs, for which there are not yet any theoretical results. Appendix~\ref{apx:QAOA_more} demonstrates parameter concentration for a range of graph-partitioning instances, indicating that parameter concentration might be a general phenomenon in QAOA. Additionally, the larger graphs show flatter loss landscapes away from significant features. This is reminiscent of expected landscapes suffering from barren plateaus.

\section{QCBM - Visualizing generalization capabilities}\label{sec:QCBM}
Generative machine learning is a popular form of unsupervised \textit{machine learning} (ML) in which the goal is to model the probability distribution that underlies an available dataset. Once trained, we can then use the model to generate similar but new data. The \textit{Quantum Circuit Born Machine} (QCBM) is a quantum generative machine learning algorithm that encodes a probability distribution over binary data in the measurement probabilities of a quantum circuit wavefunction~\cite{Benedetti2019generative}. The probability for a data sample \textbf{x} is given by the Born probability
\begin{equation}
    q_\theta(\textbf{x}) = |\langle\textbf{x}|\psi(\bm\theta)\rangle|^2,
\end{equation}
where $|\psi(\bm\theta)\rangle$ is a parameterized wavefunction that encodes the distribution $q_{\theta}(\textbf{x})$. The QCBM exploits known capabilities of quantum systems to encode complicated distributions~\cite{du2018expressive,Gao2018qml, Sweke2021learnability, gao2021enhancing}, as well as being able to generate samples from the distribution by measurement of the prepared wavefunction. However, not much is known about the practical properties of QCBMs on near to mid-term quantum devices, especially when training on realistic datasets. The most sought after practical property of any ML algorithm is related to \textit{generalization}. In the context of generative ML, generalization refers to the ability to generate new and previously unseen data that follows the true probability distribution underlying the training data. Generally in machine learning tasks, the available training data is a subset of all possible data points that would resemble the true data distribution to which one does not have access. Minimizing the training loss function or memorizing the training data does not guarantee that the trained model encodes the desired underlying distribution. This phenomenon is commonly called ``overfitting''. In other words, generalization refers to learning the true data distribution $p(\textbf{x})$ and not fitting the training data distribution $p_{\text{train}}(\textbf{x})$. 
While a rigorous investigation into the generalization capabilities of QCBMs remains an open question, we consider an empirical investigation:
we focus on similarities in the loss landscapes among datasets which resemble incomplete distributions which are sampled from the true distribution with an increasing number of samples. More specifically, we study if the loss landscape around optimized minima, when training on limited data, performs favorably on the true distribution's loss landscape. If this were not the case, one might not expect to be able to generalize from the available data.

\begin{figure}[t]
    \centering
    \includegraphics[width=1.0\linewidth]{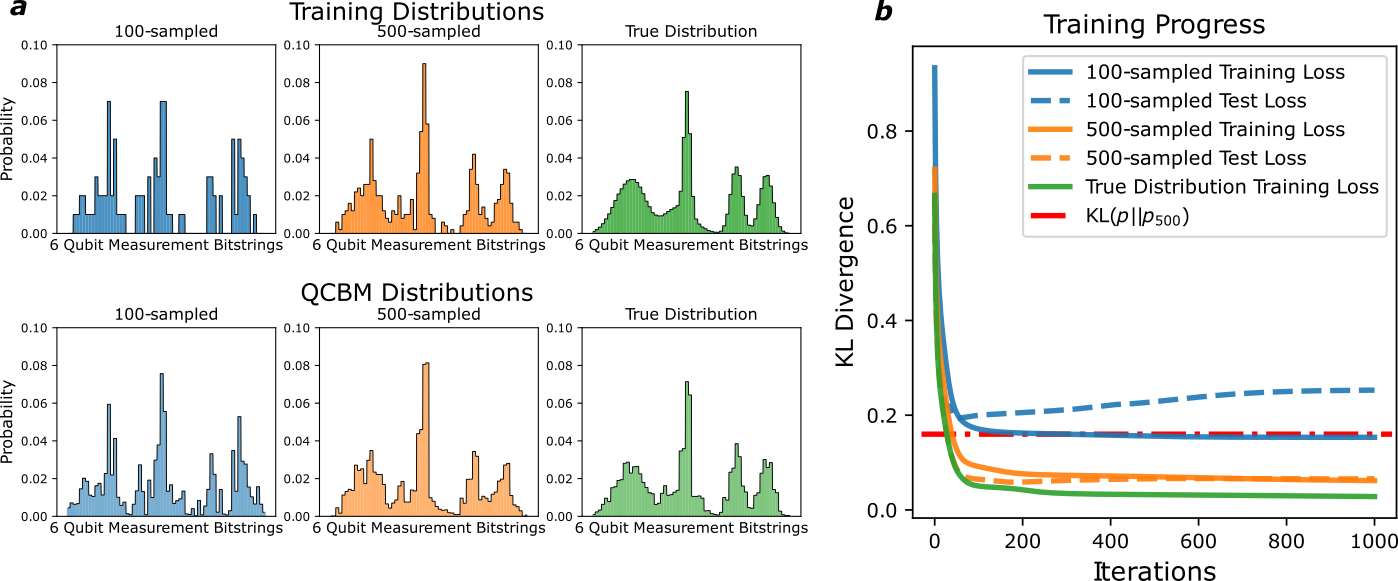}
    \caption{
    Training results of a $6$-qubit QCBM trained on three distributions: the \textit{true} data distribution $p(\textbf{x})$, as well as distributions $p_{100}(\textbf{x})$ and $p_{500}(\textbf{x})$ created by sampling the true distribution with 100 and 500 samples, respectively. All training instances were initialized with the same parameters.
    \textbf{(a)} Depiction of the training distributions and the respective final QCBM distributions. 
    \textbf{(b)} Measured KL divergence~(Eq.~\ref{eq:KL_divergence}) during training relative to the respective training distribution, as well as the cross-evaluated loss against the true distribution, which can be viewed as test loss. For the 500-sampled distribution, the training and test loss stay low. The trained models display generalization capabilities, as their losses relative to the true distribution are lower than those of their training distributions, ie. $\text{KL}\left(p||q_{\theta,100}\right) < \text{KL}\left(p||p_{100}\right)$ and $\text{KL}\left(p||q_{\theta,500}\right) < \text{KL}\left(p||p_{500}\right)$ ($\text{KL}\left(p||p_{100}\right)$ is not shown as it exceeds the y-limit). 
    }
    \label{fig:QCBM_fig}
\end{figure}

In this work, we assume explicit knowledge of the true distribution $p(\textbf{x})$ underlying the data to visualize differences in the loss landscape. The true distribution was generated by discretizing the support space of six overlapping normal distribution with random mean and variance.
We consider several sub-sampled distributions, or distributions estimated using some $k$ samples. We call such distribution a $k$-sampled distribution with distribution $p_k(\textbf{x})$. For our experiments, we constructed a 100-sampled distribution $p_{100}(\textbf{x})$ and a 500-sampled distribution $p_{500}(\textbf{x})$. The 500-sampled distribution includes the 100-sampled distribution and extends it with 400 newly drawn samples from the true distribution. These three distributions are shown in Fig.~\ref{fig:QCBM_fig}a and are chosen as training distributions for a $6$-qubit QCBM. The loss function for training here is the KL divergence:
\begin{equation}\label{eq:KL_divergence}
\begin{aligned}
    \text{KL}\left(p || q_\theta \right) & = \mathbb{E}_{\textbf{x}\sim p(\textbf{x})}\left[\log \frac{p(\textbf{x})}{q_\theta(\textbf{x})} \right]\\
    & = \sum_{\textbf{x}\sim p(\textbf{x})}p(\textbf{x})\log \frac{p(\textbf{x})}{q_\theta(\textbf{x})}
\end{aligned}
\end{equation}
Starting each respective training from the same initial parameterization, we train the QCBM for 1000 iterations of gradient descent using an exact simulator. The circuit ansatz for the QCBM is a shallow generic ansatz from Ref.~\cite{Benedetti2019generative} with only one all-to-all entangling layer surrounded by single-qubit gates.
Fig.~\ref{fig:QCBM_fig} showcases the three training distributions, the respective learned QCBM distribution, as well as a quantitative study of the training progress. In addition to the training loss, we also measure the cross-evaluated loss against the true distribution. While one does in practice not have access to the true distribution, the cross-evaluations can be viewed as the test error, in analogy to test errors in supervised learning tasks with held-back data. First, we see that the training loss is considerably lower for the 500-sampled and true distributions, indicating that smoother distributions in this representation are easier to model with the shallow circuit ansatz that is employed. The test losses follow a similar trend, however, there is a qualitative difference between the 100-sampled and 500-sampled distributions. For the 500-sampled distribution, the test loss stays close to constant with improving training loss, while for the 100-sampled distribution, an improved training loss comes at the cost of a significantly deteriorating test loss. This observation is commonly seen in classical machine learning~\cite{srivastava2014dropout}. Interestingly, the final modelled distribution by the QCBM which was trained on the 500-sampled dataset is closer in KL Divergence than the training data distribution itself, i.e. $\text{KL}\left(p||q_{\theta,500}\right) < \text{KL}\left(p||p_{500}\right)$ in Eq.~\ref{eq:KL_divergence}. This can be seen as an indication of \textit{generalization}, which is also supported by ``filling out'' certain features in the distributions shown in Fig.~\ref{fig:QCBM_fig}a. Note that the KL Divergence $\text{KL}\left(p||p_{100}\right)$ of the 100-sampled distribution compared to the true distribution is higher than the y-limit of the plot. Consequently, the QCBM model distribution $q_{\theta,100}(\textbf{x})$ displays the largest improvement in absolute terms, even though the test loss is slowly increasing.\\\\
\begin{figure}[t]
    \centering
    \includegraphics[width=0.85\linewidth]{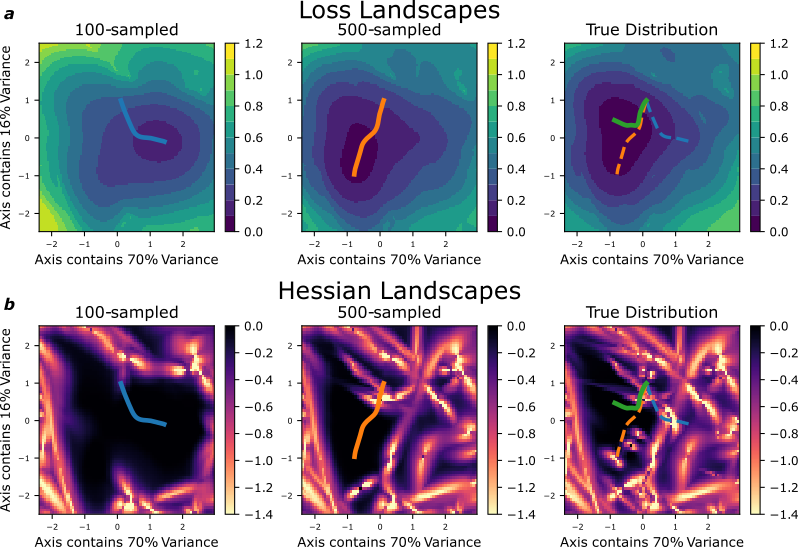}
    \caption{PCA scans on the loss landscape (\textbf{a}) and Hessian eigenvalue ratio landscape (\textbf{b}) to visualize the $6$-qubit QCBM training results in Fig.~\ref{fig:QCBM_fig}. The PCA instance is fitted to the concatenation of all three trajectories. We project the optimization trajectories on their respective landscape, as well as the cross-evaluations on the true distribution landscape. The landscapes of the 500-sampled and the true distribution are very similar, as well as parts of the trajectories. The Hessian eigenvalue-ratio scan displays more nuanced differences between the landscapes, revealing curvature which was missed by the loss scans.}
    \label{fig:QCBM_landscape}
\end{figure}
To support this quantitative study, we now apply several visualization techniques. For this purpose, we fit a \textit{Principal Component Analysis} (PCA) instance on the concatenation of all three optimization trajectories. The first two PCA components are then used as spanning directions for 2D scans on the loss landscapes. Even though PCA is used on three optimization trajectories in 39-dimensional parameter space, the first two components explain approximately 86\% of the total variance. This indicates that the optimization trajectories are each rather low-dimensional and also similar relative to each other. How they differ can be seen in Fig.~\ref{fig:QCBM_landscape} where we present PCA scans of the 100-sampled and 500-sampled distributions, as well as the true distribution. The corresponding trajectories are projected onto the scans in the colors that follow Fig.~\ref{fig:QCBM_fig}. It becomes apparent that the landscapes of the 500-sampled and the true distribution are similar, whereas the 100-sampled landscape is significantly different and has a minimum in an unfavorable region for the 500-sampled and true landscapes. Interestingly, the optimization trajectories of the 500-sampled and true distribution initially follow the same path. Then, details in the loss landscapes seem to qualitatively change the optimization outcome. In order to resolve these differences in the landscapes, we present a Hessian eigenvalue ratio plot in Fig.~\ref{fig:QCBM_landscape}b. There we plot the ratio $\frac{\lambda_{min}}{\lambda_{max}}$, where $\lambda$ are eigenvalues of the Hessian of the loss landscape at each particular point in the loss landscape. A Hessian eigenvalue-ratio scan resolves more nuanced differences between the landscapes by revealing curvature, which is missed by the loss landscape scans. The black regions where $\frac{\lambda_{min}}{\lambda_{max}} \approx 0$ indicate an extensive convex region. However, depending on the distribution, the convex region quickly turns into a highly chaotic and intricately curved landscape, as indicated by eigenvalue ratios of $\approx -1$. 

The two forms of scans depicted in Fig.~\ref{fig:QCBM_landscape} clearly show how and why each model traversed their respective landscape. The 500-sampled distribution appears to be representative enough of the true underlying distribution such that their loss landscapes are sufficiently similar for the QCBM to generalize well. Additionally, the landscapes visualize information about the \textit{width} of the minima which is difficult to study without visualization techniques. All minima lie in ``flat'' areas which in classical machine learning has been linked to good generalization ability in artificial neural networks~\cite{li2017visualizing}. Future work may study how these visualizations can be directly used for quantifying generalization capabilities.\\

\section{VQE - Comparing Ansätze}\label{sec:VQE}
The Variational Quantum Eigensolver (VQE) algorithm estimates the ground state energy of a quantum system by preparing and iteratively refining an ansatz implemented by a parameterized quantum circuit~\cite{peruzzo2014variational}.
While various ansatz designs have been proposed for different applications of VQE~\cite{wecker2015progress,kandala2017hardware,romero2019strategies,dallairedemers2019low,lee2019generalized,grimsley2019adaptive, choquette2020quantum}, there remains an open question around what makes an ansatz particularly ``good'' or effective for VQE~\cite{sim2019expressibility}.
The answer often depends on the primary goal.
For instance, if the main objective is to implement a VQE instance on a near-term quantum computer, this rules out ansätze with large circuit depths such as UCCSD~\cite{romero2019strategies} or the Variational Hamiltonian Ansatz (VHA)~\cite{wecker2015progress}. Instead, the answer in this case may lean towards more depth-efficient circuits such as Hardware-Efficient Ansatz (HEA)~\cite{kandala2017hardware}. 
In our work, we characterize and compare two VQE ansätze based on their corresponding objective landscapes for a particular problem instance.
In our case, favorable features of landscapes may include presence of high-quality minima, smoothness, convexity, and/or abundance of interspersed (high-quality) minima such that when starting from a random point, there is an efficient path to a nearby minimum.

We compare two VQE ansätze, namely the Unitary Coupled-Cluster ansatz with singles and doubles excitations (UCCSD) and its generalized variant $k$-UpCCGSD~\cite{romero2019strategies, lee2019generalized, anand2021quantum}.
The Unitary Coupled-Cluster (UCC) wave function is written as:
\begin{align}
    \ket{\psi_{\text{UCC}}} = e^{\hat{T}-\hat{T}^\dag} \ket{\phi_0},
\end{align}
where for UCCSD, the cluster operator is truncated as:
\begin{align}
    \hat{T} = \hat{T}_1 + \hat{T}_2.
\end{align}
Operators $\hat{T}_1$ and $\hat{T}_2$ are defined as:
\begin{align}
    \hat{T}_1 &= \sum_{\substack{i \in \text{occ}\\ a \in \text{virt}}} t^i_a \hat{a}^\dag_a \hat{a}_i,  \\
    \hat{T}_2 &= \sum_{\substack{i>j \in \text{occ}\\ a>b \in \text{virt}}} t^{ij}_{ab} \hat{a}_a^\dag \hat{a}_b^\dag \hat{a}_j \hat{a}_i,
\end{align}
where the former generates single excitations, and the latter generates double excitations. Here, $\hat{a}^\dag$ and $\hat{a}$ correspond to fermionic creation and annihilation operations, respectively, and $i,j$ and $a, b$ index occupied and virtual orbitals, respectively. 
The $k$-UpCCGSD ansatz comprises $k$ layers/repetitions of generalized singles and paired double excitations.
Unlike in UCCSD, generalized operators additionally allow excitations in the occupied-occupied and virtual-virtual subspaces.
A generalized paired double excitation operator has the form $\hat{a}_{p\uparrow}^\dag \hat{a}_{p\downarrow}^\dag
\hat{a}_{q\downarrow} \hat{a}_{q\uparrow}$, where indices $p, q$ run over occupied \textit{and} virtual orbitals.
For more information on the structures of these UCC ansätze, we refer the readers to Refs.~\cite{lee2019generalized, anand2021quantum}.

Performance of the two ansätze has been compared in terms of their effectiveness in estimating the ground state energies of various molecular systems as well as their circuit resources such as parameter count and circuit depth \cite{lee2019generalized, grimsley2020trotterized, sim2020adaptive}. 
We consider the molecular system $H_4$ with $8$ qubits, which has been previously considered as a benchmark system for VQE ansätze \cite{romero2019strategies, lee2019generalized} as well as multireference methods in quantum chemistry.
Here, we extend the comparisons in previous works by not only tracking the quality of local minima but also visualizing the position of the minima and the underlying loss landscape.
We consider the rectangular configuration as shown in Fig.~$4$ of Ref.~\cite{romero2019strategies}, in which $r$ is scanned across a range while $d$ is fixed at $1.23\AA$. As noted in Ref.~\cite{lee2019generalized}, when $r=1.23\AA$, and the molecular system assumes a square geometry, this problem becomes challenging for single-reference methods such as UCCSD. The authors showed that $k$-UpCCGSD performed well for estimating the ground state energies for this system using small values of $k$, or the number of repetitions. In the following, we reproduce and extend their results using visualization techniques to further highlight the strengths of this ansatz.
The VQE simulations were implemented using Tequila \cite{kottmann2021tequila} and Qulacs \cite{Suzuki2021Qulacs} packages. 

\begin{figure}[t]
    \centering
    \includegraphics[width=0.85\linewidth]{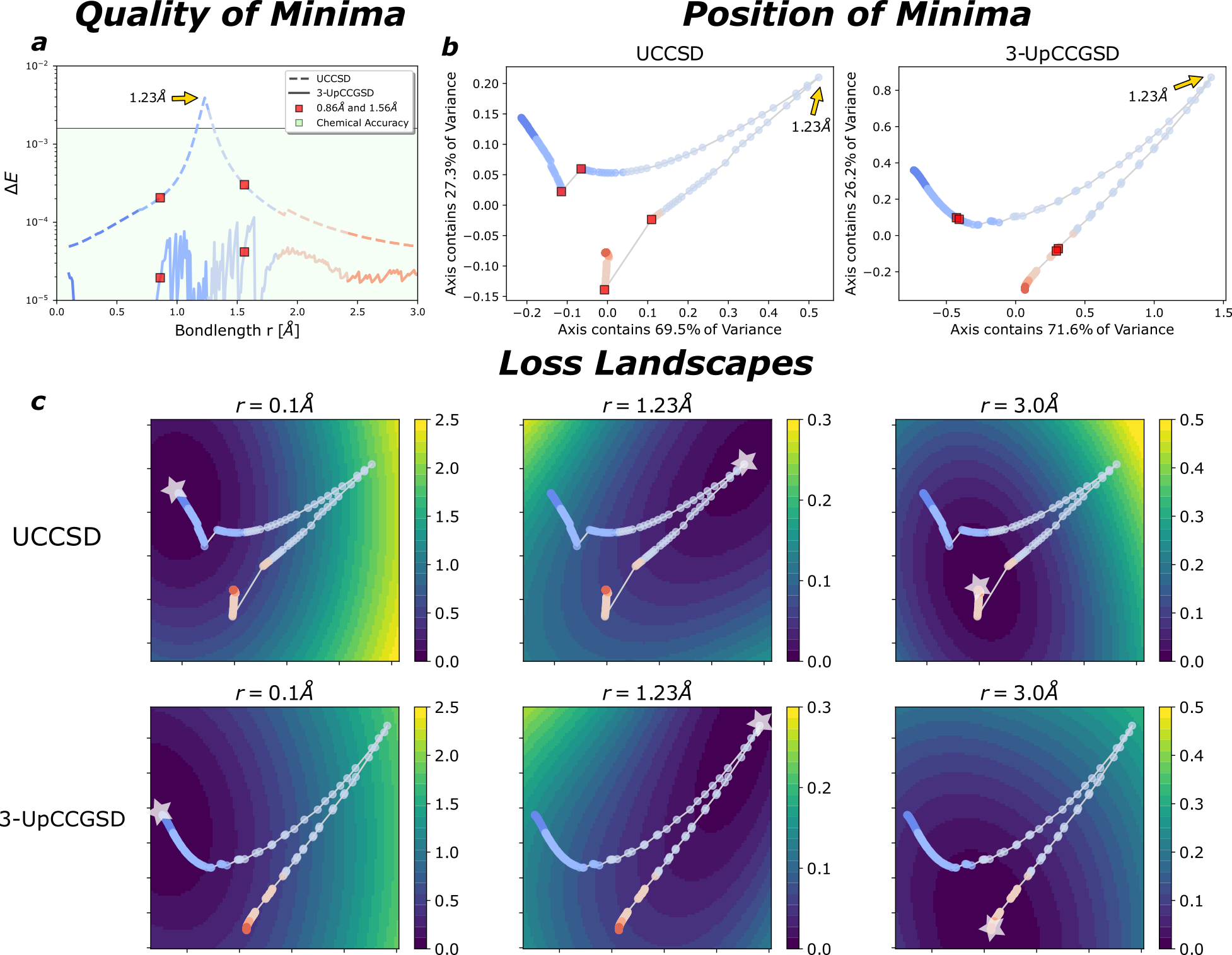}
    \caption{Comparison of the UCCSD and $k$-UpCCGSD ($k=3$) ans\"atze on the $H_4$ molecular system with $8$ qubits at different bond lengths $r$. \textbf{(a)} Energy error $\Delta E$ for both ansätze and indication of chemical accuracy as $\Delta E \approx 0.00159$ Ha. 
    UCCSD is initialized close to zero, in which parameter values of zero corresponds to the mean-field solution of the Hamiltonian. For $3$-UpCCGSD, the solution at $r=0.1\AA$ is found from a random initialization $\in[-\pi/2,\pi/2]^{\otimes 39}$, while minima for larger $r$ are found by initializing with the solution at the previous $r$. \textbf{(b)} PCA projection of the collection of minima at all $r$. The corresponding bond lengths are encoded in the color relative to panel \textbf{a}. \textbf{(c)} Scanning the energy landscape corresponding to $r=0.1\AA$, $r=1.23\AA$ and $r=3.0\AA$, respectively. The minima appear wider than the total distance between all minima.}
    \label{fig:VQE_minima_path}
\end{figure}

To initialize parameters of UCCSD, we employed MP2 amplitudes though we additionally observed that initializing each parameter using any near-zero value also performed well and led to the same minimum as when employing MP2 amplitudes~\cite{romero2019strategies}. Random initial parameters, e.g. in the range $[-\pi/2, \pi/2]$, did not converge to a local minimum below chemical accuracy.
To initialize $k$-UpCCGSD, we observed that the near-zero initialization technique employed for UCCSD did not perform well. Instead, we randomly initialized each parameter from the range $[-\pi/2, \pi/2]$, where the ansatz typically converged well.

Fig.~\ref{fig:VQE_minima_path}a yields similar results as Ref.~\cite{lee2019generalized}, in which 3-UpCCGSD estimated the ground state energies across various geometries (varying $r$) to sufficient accuracy. Meanwhile, as expected, UCCSD fails to achieve the same accuracy, especially in regions near and at the square geometry (indicated on the plot using red square markers and the yellow arrow).
We extended the analysis by taking the final/optimized parameter values of each ansatz at all considered geometries and performed PCA on these parameter vectors, as shown in Fig.~\ref{fig:VQE_minima_path}b. For UCCSD, there is a jump or discontinuity in two regions of the projected landscape of optimized parameters, which are regions leading up to the square geometry. At $r=d=1.23\AA$, we observe a sharp turn or a cusp in the optimized parameters.
Minima immediately preceding and following the square geometry are close in distance as their corresponding geometries form similar rectangular configurations i.e. their sides are proportional. 

That is, the side ratios $\frac{1.23-\epsilon}{1.23} \approx \frac{1.23}{1.23+\epsilon}$ for small values of $\epsilon$.
This likely causes the rapid turn in the parameters of minima after $r=1.23\AA$.
As $\epsilon$ increases, the geometries (and their corresponding minima) diverge.
For 3-UpCCGSD, the overall landscape of optimized parameters are similar, though the discontinuities observed in UCCSD are removed for 3-UpCCGSD. In fact, the minima for 3-UpCCGSD that are shown in Fig.~\ref{fig:VQE_minima_path} were found using successive initialization, in which the minimum at the $i+1$-th bond length is initialized with the final parameter values of the $i$-th bond length. This appears to imply that the ansatz is sufficiently flexible to capture the change in the ground state wavefunction, especially near the square geometry. 
Another distinction between the two ans\"atze is the characteristics of minima. For UCCSD, various initialization techniques, including MP2 amplitudes, near-zero values, and successive initialization, led to the same minima. This implies relatively large widths of minima or a degree of robustness to parameter initialization (near zero). 
For $3$-UpCCGSD, random parameter initializations from $[-\pi/2, \pi/2]$ yielded different optimization trajectories and final parameters, implying the presence of a number of (local) minima. As can be seen in Fig.~\ref{fig:VQE_minima_path}c, the minima in this case are wide enough such that one can initialize this model at any of the previously found solutions to optimize for a different bond length. However, previous work has shown that, for larger molecules, one may require up to hundreds of parameter initializations to converge to good local minima, as well as larger values of $k$~\cite{lee2019generalized}. The underlying reason is that, while the ansatz is sufficiently powerful, it does not have a natural initialization strategy like UCCSD. For $H_4$, we can begin to study this property of $k$-UpCCGSD by applying the Nudged Elastic Band Algorithm (NEB, see Secs.~\ref{sec:example},\ref{apx:elastic_band}).
\begin{figure}[t]
    \centering
    \includegraphics[width=1\linewidth]{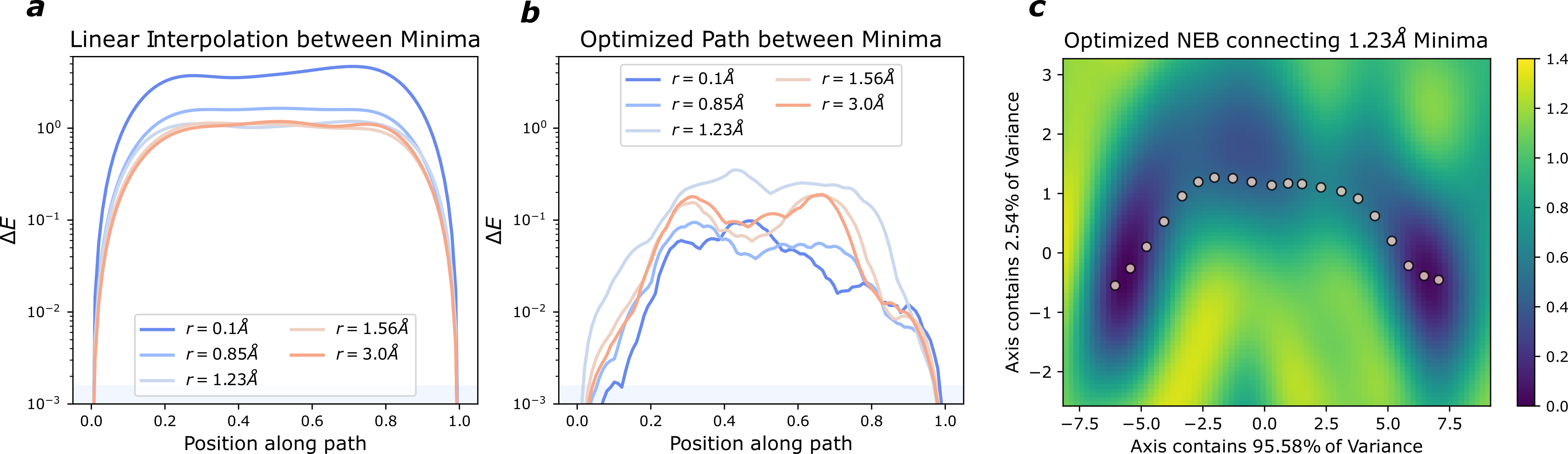}
    \caption{Demonstration of the \textit{Nudged Elastic Band} algorithm applied on pairs of minima found with the $k$-UpCCGSD ansatz ($k=3$) at different molecular distances $r$. \textbf{(a)} 1D scan of the loss landscape by linear interpolation between the parameters of the respective minima. The minima at $x=0$ are the same minima as in Fig.~\ref{fig:VQE_minima_path}, while the minima at $x=1$ stem from a second random initialization on $r=0.1\AA$ and successive initialization at larger $r$ with the solutions found previously. As indicated the light-green shaded area, all minima achieve chemical accuracy. \textbf{(b)} Energy scanned along the optimized path with the Nudged Elastic Band. \textbf{(c)} Depiction of a PCA scan on the optimized path on the $r=1.23\AA$ landscape. It is apparent that the minima are ``isolated'' in that they are not connected by a path of very low energy (i.e. compared to chemical accuracy or $10^{-3}$ Ha).
    However, any random initialization of parameters seems to result in a good minimum, indicating a generally favorable landscape with many (isolated) high-quality local minima.}
    \label{fig:VQE_NEB}
\end{figure}
To gain a better understanding of the landscape of $k$-UpCCGSD, we apply the NEB algorithm on pairs of minima at various bond lengths, as shown in Fig.~\ref{fig:VQE_NEB}.  The second collection of minima was found starting from a different random initialization.
We observe that at all considered bond lengths, the optimized energy path connecting two minima has a relatively high energy barriers, significantly higher than $10^{-3}$ Ha or chemical accuracy.
This observation that the local minima are rather disconnected is consistent with previous results and may explain the difficulty in convergence behavior of the ansatz for larger molecules. Interestingly, our visualizations of the continuously stretched $H_4$ molecule suggest one may use a solution at a different bond length, perhaps one that is easier to optimize, and apply those solution parameters as initial parameters for the molecular geometry of interest.

To conclude, we show that the two types of UCC ansätze, UCCSD and $k$-UpCCGSD, have significantly different properties. UCCSD has a natural parameter initialization scheme for a given problem (MP2 or zero initialization) but may suffer from lacking expressivity in describing strongly correlated molecular systems. On the other hand, $k$-UpCCGSD is potentially a highly flexible and expressive ansatz, which we further highlight by continuously manipulating the ansatz parameters to accurately estimate the ground state energy of nearby geometries. However, this comes at the cost of lacking a clear strategy for initial parameters, as well as having high energy barriers between minima in the energy landscape. 
Hence, the ansatz faces a different type of challenge when solving for larger molecules.

\section{Noise - Effects of finite-sampling and stochastic optimization}\label{sec:NOISE}
In practice, evaluations of objectives and gradients in near-term quantum algorithms are estimated using a finite number of samples or \emph{shots} from the quantum computer. While there are other sources of noise, we restrict our analysis to the noise from finite-sampling, namely its effect on the optimization landscapes of VQAs.

We investigate the effects of finite-sampling on the optimization and landscapes of $4$-qubit Quantum Circuit Born Machine (QCBM, see Sec.~\ref{sec:QCBM}) and Variational Quantum Eigensolver (VQE, see Sec.~\ref{sec:VQE}) instances.
In the limit of infinite number of samples, the landscapes are smooth, in which we can resolve even delicate features. This has been the case in all sections so far in this work. 
Depending on the type of loss function, using a finite number of shots can greatly impact the loss landscape.

First, we compare the effect of finite-shots on two common optimizers aiming to train a QCBM on a random four-bit distribution. The optimizers are \textit{Simultaneous Perturbation Stochastic Approximation} (SPSA)~\cite{Spall1992SPSA} and \textit{Covariance Matrix Adaptation Evolutionary Strategy} (CMA-ES)~\cite{hansen2016cma}. SPSA approximates the local gradient of the loss by calculating a numerical gradient in a random direction. CMA-ES is a gradient-free optimizer which evolves a population of parameter vectors according to an estimation of the local covariance matrix. Commonly in SPSA, one only estimates the gradient in a single random direction while CMA-ES uses a larger population in the evolutionary strategy. 
For a fair comparison, we allot the same number of function evaluations for both optimizers, i.e. we initialize CMA-ES with a population of six and for SPSA, we average the gradient over three random directions with two evaluations per direction. Additionally, we start all optimizations from the same initial parameters and set the learning rate of SPSA and initial step size of CMA-ES to $0.1$. Note, that for CMA-ES, this indicates the standard deviation of the initial perturbations used to evolve the population, not a static step size.
\begin{figure}[t]
    \centering
    \includegraphics[width=0.85\linewidth]{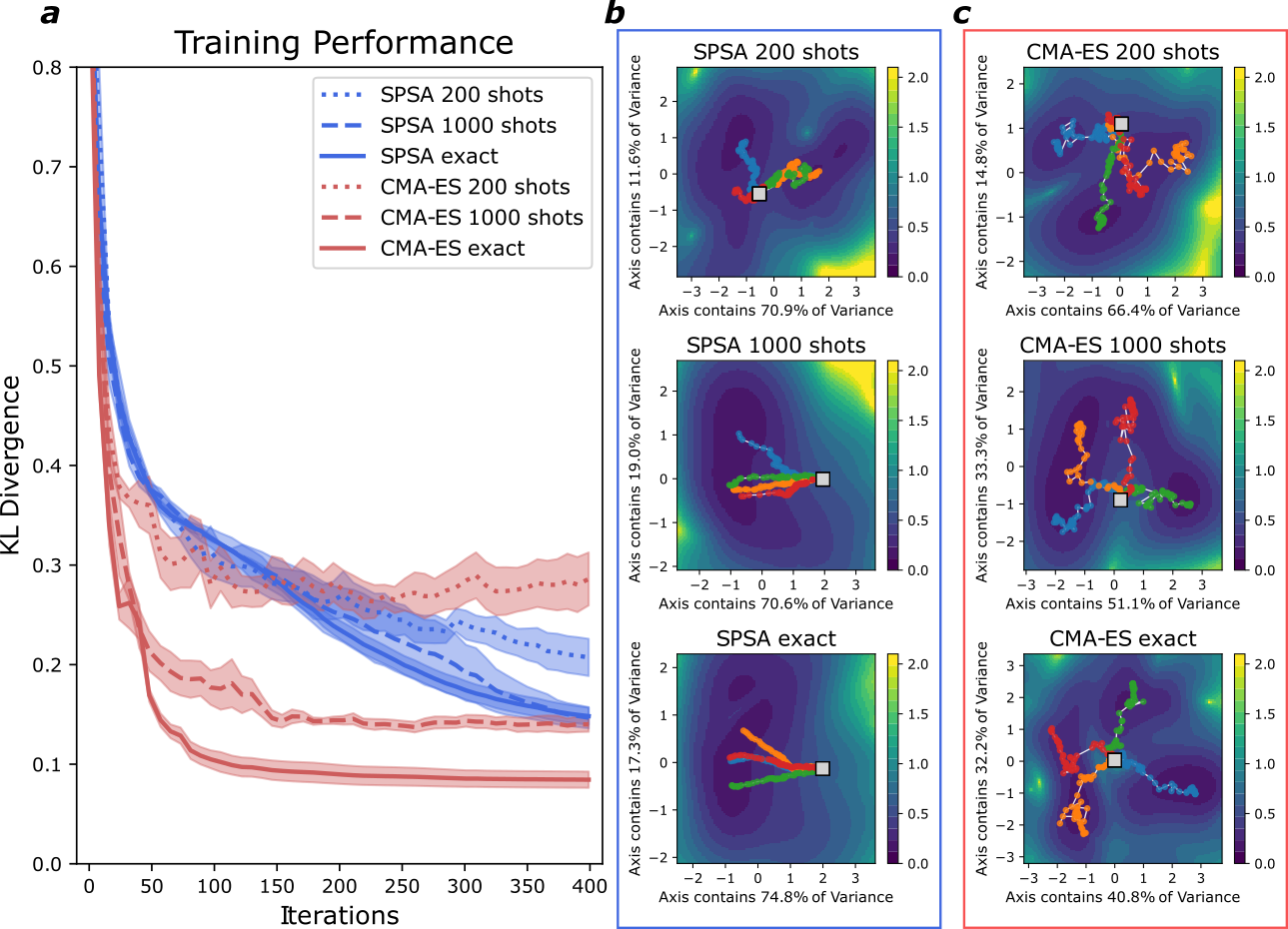}
    \caption{\textbf{(a)} Average KL divergence values during training of a $4$-qubit QCBM with SPSA optimizer and the CMA-ES optimizer, as well as variance over four re-starts, when using $200, 1000$ measurements (shots) per loss function evaluation, or the exact circuit wavefunction. All training runs were started from the same initial parameters and are allowed six loss function evaluation per optimization step. \textbf{(b},\textbf{c)} PCA scans performed on all optimization trajectories with SPSA (\textbf{b}) and CMA-ES (\textbf{c}), and projection of the respective trajectories on top of the scanned loss landscape. While the SPSA gradient estimates become more accurate and consistent with a larger number of shots, the gradient-free CMA-ES solver shows a more non-local optimization and finds several good minima from the same initialization.}
    \label{fig:NOISE_optimizer}
\end{figure}
Fig.~\ref{fig:NOISE_optimizer}a displays a quantitative comparison between the optimizers over four repetitions. It appears that CMA-ES outperforms the stochastic gradients when allowed a large number of shots; however, it seems less robust to the sampling noise. In contrast, SPSA performs approximately equally with a low number of shots or the exact wavefunction. This indicates that the stochastic gradient evaluation of SPSA in random directions is a large error source and potentially larger than noise induced by finite sampling. Visualizing these results in Fig.~\ref{fig:NOISE_optimizer}b clearly shows the different characteristics of the two optimizers. While the four SPSA trajectories become more deterministic and one-dimensional, CMA-ES becomes less one-dimensional and explores a larger space. Even though we study the optimizers using an exact simulator, these visualizations do not clearly depict how the loss landscape is affected by finite-sampling. In the following, we study this more explicitly.
\begin{figure}
    \centering
    \includegraphics[width=0.99\linewidth]{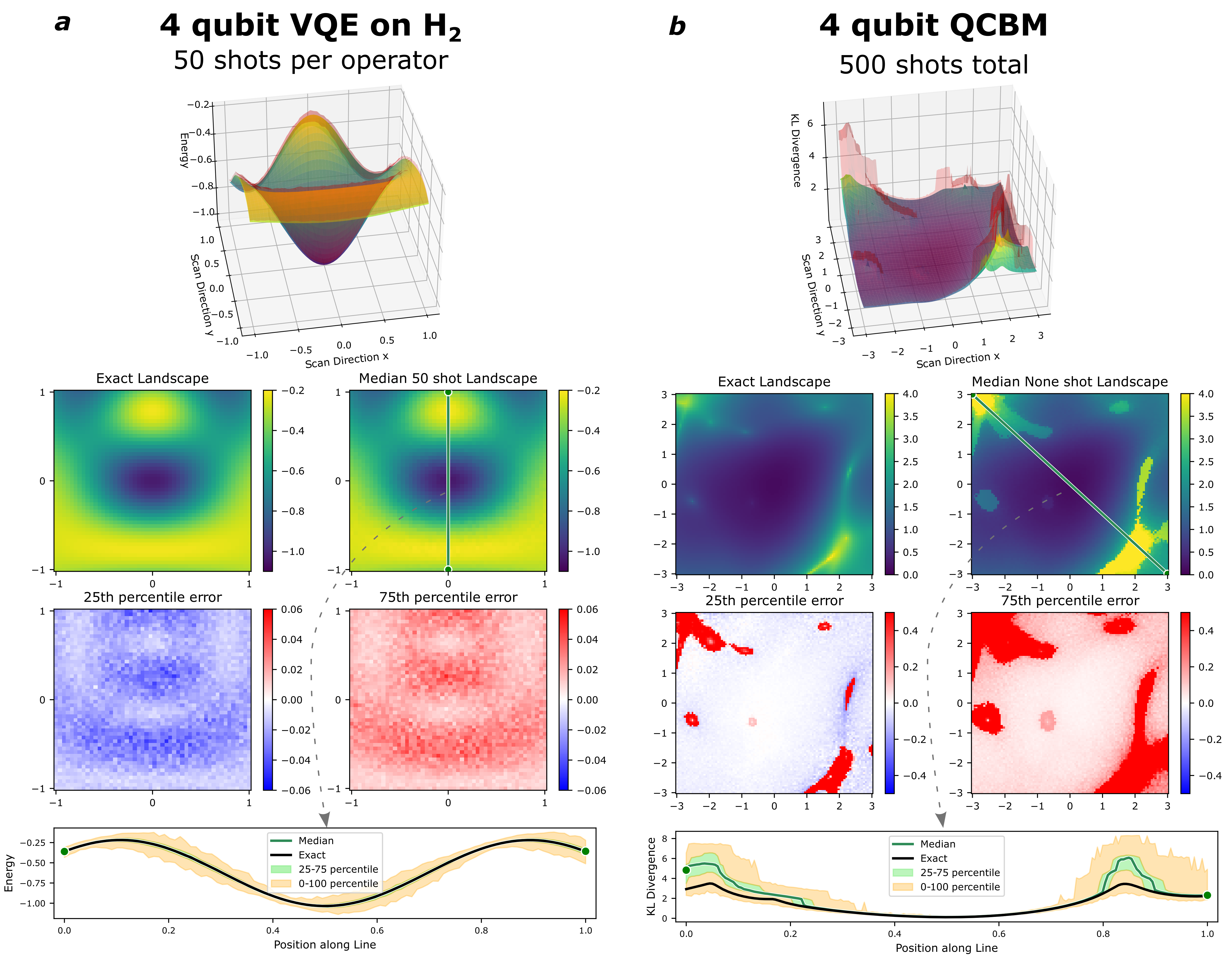}
    \caption{Showcase of the effect of finite-sampling for a $4$-qubit Hydrogen molecule ($H_2$) VQE instance (\textbf{a}), as well as $4$-qubit QCBM trained on a random distribution (\textbf{b}). The VQE task minimizes energy of a variational quantum state on a Hamiltonian, while the QCBM is trained on the KL divergence against the training data distribution. The exact loss landscapes are compared against the median loss landscape with finite shots, as well as the 25th and 75th percentile errors relative to the exact values. Statistics are taken from repeating the 2D scans 100 times and the 1D scan 1000 times. For the VQE instance, we take 50 shots to evaluate each operator in the Hamiltonian while for the QCBM instance, we take a comparable 500 shots to estimate the current distribution and calculate the KL divergence. The variance in the VQE energy estimation is symmetric around the exact expectation, while the KL divergence estimation with a QCBM  is biased upwards.}
    \label{fig:NOISE_shots}
\end{figure}
We now compare the noisy landscapes of a $4$-qubit VQE instance for the $H_2$ molecule and a $4$-qubit QCBM instance trained with KL divergence on a random distribution. The VQE instance uses a UCCSD ansatz with only two parameters, while the QCBM is implemented with a hardware-efficient ansatz and 14 parameters. For the VQE instance, we use 50 shots to evaluate each operator in the Hamiltonian, and 500 samples to estimate the current distribution in the QCBM and calculate the KL divergence. These numbers are chosen such that, depending on how operators in the same basis of the VQE Hamiltonian are grouped, the numbers stay comparable. Both landscapes are studied around a previously found minimum. Fig.~\ref{fig:NOISE_shots} shows the landscapes from exact simulation against the median loss landscapes using a finite number of samples, as well as the 25th and 75th percentile errors relative to the exact values. Statistics are taken from repeating the 2D scans 100 times. The 1D scans present a cross-section of the 2D scans and are performed along the indicated linear paths, using statistics of 1000 noisy repetitions.

For evaluation of the energy expectation in VQE, the median landscape is very close to the exact landscape. The variation is symmetric around the exact value (average and median are the same), though not uniform across all points in the landscape. This is generally expected when measuring expectations of quantum operators, where the number of shots $n_s$ for a given uncertainty $\epsilon$ in the estimation scales as $O(\frac{1}{\epsilon^2})$. In contrast, for the QCBM loss function, the median landscape is very different from the exact landscape. It becomes clear that finite-sampling biases the evaluation of the KL divergence upwards. In different regions of the parameter space, even low-percentile values lie above the exact loss. The reason for these observations lies in the singular nature of the KL divergence~(see Eq.~\ref{eq:KL_divergence}). If a measurement outcome $\textbf{x}$ is not observed, i.e., the experimental value is $q_\theta(\textbf{x})=0$,  but \textbf{x} has a non-zero probability $p_{train}(\textbf{x})$ in the training data distribution, then the model probability $q_\theta(\textbf{x})$ is set to $\epsilon$ (with $\epsilon=10^{-6}$) such that the cost function term $p_{train}(\textbf{x})\cdot\log(q_\theta(\textbf{x}))$ avoids the numerical singularity of the logarithm. This is less likely when close to the minimum, when samples are drawn from a distribution which is closer to the underlying data distribution $p_{data}$ from which the training data $p_{train}$ itself was drawn. However, it is likely far from a local minimum, where artifacts arising from singularities can lead to connecting local maxima and effectively creating barriers in the loss landscape.

For this reason, recent work has investigated training QCBMs as implicit generative models where the training loss can be estimated from samples more robustly~\cite{mohamed2017implicit,leadbeater2021fDivergence}.

On the other hand, while VQE appears to be less systematically affected by finite-samples simulation, the variance in estimation of the energy poses a different problem. VQE instances require very accurate energy estimates, usually below chemical accuracy of roughly $\approx 10^{-3}\text{Ha}$, and also high confidence in that estimate. While our results indicate that training far away from the minimum may be robust even with few shots, in practice one requires a very large number of shots to estimate energies to chemical accuracy~\cite{gonthier2020identifying}.

The visualization techniques assist us in understanding and appreciating fundamental differences between these two VQA algorithm, their respective loss functions, and how shot noise affects each of them.

\newpage
\section{Summary and Outlook}
In this work, we have reviewed various visualization techniques and presented their corresponding code from
the open-source Python library \textit{orqviz}. 
We first presented the techniques and applied them to investigate a classical toy loss function in Sec.~\ref{sec:example}. To demonstrate the utility of \textit{orqviz} for quantum computing, we similarly applied these techniques for a range of VQAs. 
In Sec.~\ref{sec:QAOA}, we studied the phenomenon of \textit{parameter concentration} in QAOA, which arises as a natural application for visualization. A wide range of QAOA instances could be tested in this way, where more theoretical work is difficult. 
Sec.~\ref{sec:QCBM} showcased how the loss landscape of a QCBM depends on the training dataset which is, in practice, a subset of the true underlying data distribution. In this context, visualization was shown to be an interesting approach for studying the generalization and training capabilities of quantum generative models under the realistic assumption of limited data. 
In Sec.~\ref{sec:VQE}, we compared the UCCSD and $k$-UpCCGSD ansätze for the VQE on the $H_4$ molecular system and highlighted fundamental differences between the two ansätze. Particularly promising is the application of the \textit{Nudged Elastic Band} to study connectivity of local minima which may be linked to trainability of VQAs.
Finally, in Sec.~\ref{sec:NOISE}, we inspected the effects of finite-sampling noise in VQE and QCBM loss landscapes. Our results revealed robustness in estimation of an operator expectation and instabilities of singular loss functions such as the KL divergence.

Beyond VQAs discussed in this work, visualization techniques in \textit{orqviz} can be readily applied to other parameterized models in quantum computing and adjacent fields.
A natural extension to this work includes investigation of changes in VQA landscapes with error-mitigation~\cite{giurgica-tiron2020digital} or amplitude estimation techniques~\cite{wang2021minimizing}. Similarly to the \textit{parameter concentration} phenomenon studied in Sec.~\ref{sec:QAOA}, one might be able to visually demonstrate phenomena which have been investigated theoretically and numerically, consequently gaining further understanding which might inspire the development of new algorithmic choices. Visualization can be used to examine \textit{barren plateaus}~\cite{mcclean2018barren}, landscape features such as \textit{narrow gorges}~\cite{arrasmith2021equivalence}, how different loss function properties affect the optimization landscape (e.g., Refs.~\cite{cerezo2020variational, kieferova2021divergences}), and other important aspects to consider when training VQAs.

There also has been an increasing amount of work in recent years on identifying differences between quantum and classical models (e.g., Refs.~\cite{Gao2018qml,glasser2019expressive, Sweke2021learnability, gao2021enhancing, anschuetz2021critical}). Visualization may be another approach to compare classes of parameterized models for a range of problem instances in terms of, for example, the quality, number, or connectivity of local minima. This extends to quantum-classical algorithms where the aim is to enhance classical algorithms with a quantum component~\cite{perdomo2017opportunities,benedetti2017helmholtz,khoshaman2018QVAE, anschuetz2019associate,rudolph2020generation}, as well as classical algorithms for physics, such as variational machine learning models for quantum states~\cite{carleo2017solving, jonsson2018neural, Melko2020MLQuantumStates}. In this case, one could use the visualization techniques to evaluate the quality of their approximation to the true quantum state.
Outside of strictly research domains, visualization can be useful for educational purposes. Not only are these techniques helpful for less experienced researchers to understand concepts in the current field of quantum algorithms, but experts can also explore problems from a new perspective. 

We expect visualization techniques to significantly contribute to the development and analysis of algorithms that employ parameterized quantum models. As demonstrated in this work, the capabilities provided by \textit{orqviz} are flexible enough to deepen our understanding of VQAs as well as play a critical role in the design of parameterized quantum algorithms. 
As such, we invite readers to use and contribute to the \textit{orqviz} Github repository: \href{https://github.com/zapatacomputing/orqviz}{\textbf{github.com/zapatacomputing/orqviz}}.

\begin{acknowledgments}
The authors would like to acknowledge Dax Koh, Marta Mauri, Peter Johnson and Brian Dellabetta for their feedback on the manuscript. The authors would like to acknowledge the Orquestra\texttrademark~platform by Zapata Computing Inc. for generating the data presented throughout this work. A.R. acknowledges Zapata Computing for hosting his internship. E.R.A. is supported by the National Science Foundation Graduate Research Fellowship Program under Grant No. 4000063445.
\end{acknowledgments} 

\newpage
\bibliographystyle{unsrt}

\appendix

\section{Computational resources}\label{apx:comp_resources}
To better understand the trade-off between applying the visualization techniques and performing more quantitative studies, e.g. running more optimizations, we briefly discuss the computational resources required to effectively apply the visualization techniques. For details on the techniques, we refer to Sec.~\ref{sec:example} \& Appendix~\ref{apx:viz_techniques}.

Of all the techniques reviewed in this work, 1D scans are the least computationally intensive technique. For instance, a 1D scan with 100 evaluations along a path requires as many loss function evaluation as two gradient descent steps using numerical gradients on a 25-parameter VQA. Consequently, one could perform dozens of 1D scans at the computational cost of one full optimization run with gradient descent.

The 2D scans are more informative than 1D scans, but they require resources that scale quadratically with the desired resolution. For example, a $30\times30$ (900 evaluations) grid scan corresponds an equivalent of 18 gradient descent steps; a $40\times40$ (1600 evaluations) grid 2D scan to 32 gradient descent steps. While certainly more expensive, 2D scans are highly flexible and can be used to visualize results from PCA, computed Hessians, and the NEB algorithm.

By contrast, a single Hessian evaluation scales quadratically with the number of parameters, such that an $n$-parameter model requires evaluation of $\frac{n\cdot (n-1)}{2}$ matrix entries; each requiring $4$ loss function evaluations in typical cases. A particularly intensive calculation is that of the Hessian eigenvalue ratio 2D scan in Fig.~\ref{fig:QCBM_landscape}b, which estimates the Hessian on a 2D grid of parameters.

Finally, the NEB algorithm, while certainly resource intensive, is promising for its exploration of the loss landscape beyond linear scans. It can be used to study width and connectivity of minima in VQA applications, as well as to find saddle points between points in high-dimensional space. 

While visualization techniques can be expensive for larger simulations, their purpose is not to solve a particular problem but to provide insights on the nature of the problem, e.g. the difficulty of the problem through the perspective of its landscape characteristics. Often the problem sizes we can consider are limited, but insights from these small cases are nevertheless useful for a better understanding of especially heuristic algorithms. Modern quantum simulators, like the \textit{Qulacs} simulator~\cite{Suzuki2021Qulacs} used throughout this work, are exceptionally fast and optimized for low number of qubits such that a single 2D scan can be generated in a few seconds. When limited by computational resources, we particularly recommend PCA scans, as applying the PCA algorithm to parameter trajectories does not require additional quantum resources, and the accompanying 2D scans are inexpensive compared to evaluating the Hessian or applying the NEB algorithm.

\section{Details of Visualization Techniques}\label{apx:viz_techniques}

\subsection{Parameter symmetry and normalizing distances}\label{apx:periodic_wrap}
In classical artificial neural networks, it has been demonstrated that loss landscape visualization results can be misleading if symmetries in the parameterized model are not considered~\cite{li2017visualizing}. For example, if a network uses the popular ReLU activations, the output of the network is preserved when multiplying a layer's weights by an arbitrary factor and the following layer is divided by the same factor. Interpolating between parameter vectors thus traverses a distorted loss space, and one cannot confidently choose scales at which to apply the visualization scans. This problem can be circumvented by normalizing network layers to a common order of magnitude.

Quantum circuits also contain symmetries, most prominently in the periodicity of the applied quantum gates. For example, most operations are defined to be $2\pi$-periodic with respect to their parameterization (note that conventions for gates can vary). This means that there exist infinitely many ``copies'' of a given parameter vector which differ by a multiple of $2\pi$ in each entry of the vector. One might expect that by normalizing parameters to a fixed domain, e.g. $[0,2\pi)$ or $[-\pi,\pi)$, this symmetry is accounted for. However, when calculating differences between parameter vectors, this is not sufficient, as the shortest path between two parameter vectors is between the closest copies of those parameter vectors. For example, the linear path $0.1 \longrightarrow (2\pi-0.1)$ can be traversed as $0.1 \longrightarrow -0.1$  and has a length of $0.2$.\\\\
\textit{orqviz} offers the utility function ``\textit{relative\_periodic\_wrap}'' which wraps a parameter vectors to their closest copy relative to a reference parameter vector.
\begin{samepage}
\begin{mintedbox}{python}
import numpy
from orqviz.geometric import relative_periodic_wrap

parameter_vector_1 = numpy.array([0.1, 1.2])
parameter_vector_2 = numpy.array([2*numpy.pi - 0.1, 1.3])
wrapped_parameter_vector2 = relative_periodic_wrap(parameter_vector_1, parameter_vector_2, period=2*numpy.pi)

# wrapped_parameter_vector2 = [-0.1, 1.3]
\end{mintedbox}
\end{samepage}
\begin{figure}[t]
    \centering
    \includegraphics[width=0.8\linewidth]{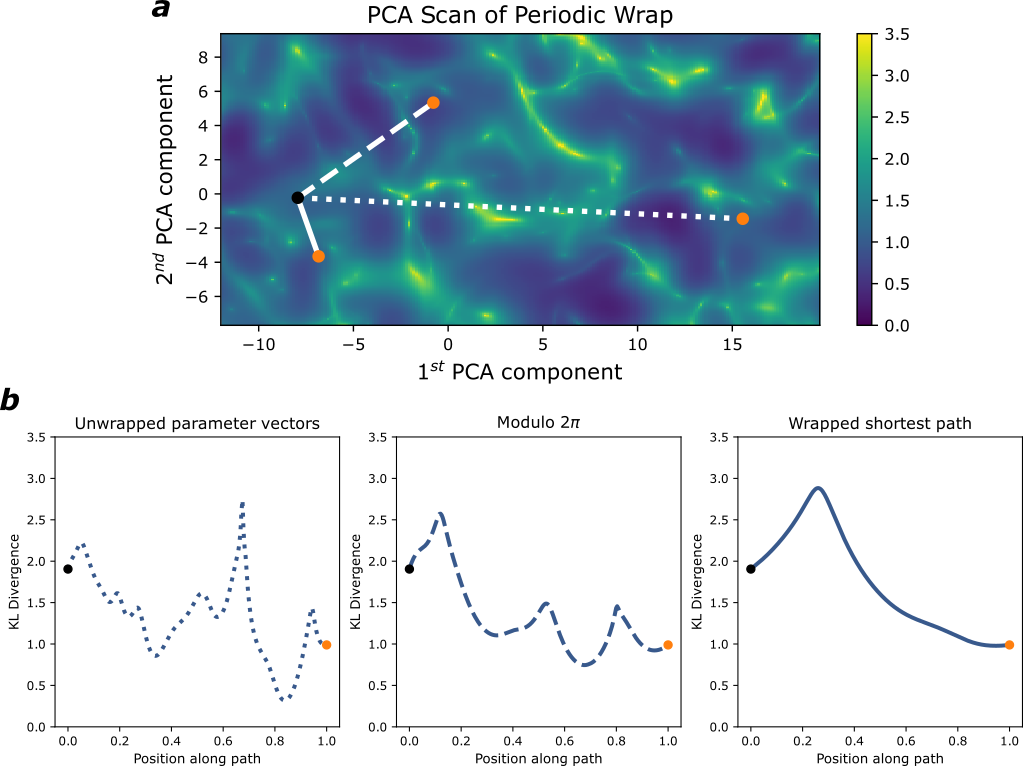}
    \caption{Demonstration of the effect of respecting symmetries in VQA applications. We scan linear interpolation trajectories between two points (black and orange) in $14$-dimensional parameter space with $2\pi$ symmetry. The three copies of the orange point are equivalent and have the same loss. The differences are that the parameters in the parameter vector are $\in[0,4\pi]^{\otimes 14}$ (dotted lines), $\in[0,2\pi]^{\otimes 14}$ (dashed lines) and the closest wrapped copy of the point relative to the black point (solid line). The respective helper function ``\textit{relative\_periodic\_wrap}'' is available with the \textit{orqviz} library.
    \textbf{(a)}~PCA scan on the three trajectories in parameter space. The scan reveals relative distances between the three copies of the same orange point. \textbf{(b)}~1D interpolation scans of the loss function between along the three linear interpolation trajectories. The wrapped point is closest to the black point and thus the loss landscape has lower frequency features.}
    \label{fig:periodic_wrap}
\end{figure}
To visualize these symmetries, Fig.~\ref{fig:periodic_wrap} shows an example of interpolating between two random points (black and orange) randomly sampled in the domain $[0,4\pi]^{\otimes14}$. The loss landscape is that of a $4$-qubit QCBM with the KL divergence (Eq.~\ref{eq:KL_divergence}) as the loss function. The three orange points are equivalent as they only differ by multiples of the $2\pi$ parameter period. The dotted line indicates the linear interpolation between the original points in the $4\pi$ domain along the real number line. The dashed line interpolates between the same points modulo $2\pi$, to restrict the entries of the parameter vectors to $\in[0,2\pi]^{\otimes14}$. This does not affect the losses of the black and orange point. Finally, we demonstrate the ``\textit{relative\_periodic\_wrap}'' function in \textit{orqviz} to wrap the orange point to its closest copy relative to the black point (solid line). The loss observed via linear interpolation between the parameter vectors varies vastly between the different paths taken. A PCA scan on these linear paths approximately reveals the difference in relative distance between the points, as well as the loss landscape.

\subsection{Fundamental Scans}\label{apx:simple_scans}
Many visualization techniques presented in this work involve \textit{scanning} the loss landscape. \textit{Scanning} refers to evaluating the loss function $\mathcal{L}$ on manifold $\mathcal{M}\equiv\{\bm\theta\}_{\bm\theta\in\mathcal{M}}$ in parameter space. Visualizing the loss landscape then consists of choosing the manifold $\mathcal{M}$ and how to plot the function $f\left(\mathcal{M}\right)=\{\mathcal{L}\left(\bm\theta\right)\}_{\bm\theta\in\mathcal{M}}$. We encourage practitioners to be creative with how to utilize these tools and how to combine them to build and present their work.
\subsubsection*{1D scans}
1D scans are a light-weight technique for studying loss function landscapes. Generally, one can evaluate the loss $\mathcal{L}$ (or another metric of interest) along a path $p$:
\begin{equation}
    f(t) = \mathcal{L}\left(p(t)\right)
\end{equation}
where $t$ is the path parameter, for example $t\in[0,1]$.\\
One choice is implementing a \textit{linear} scan along a random direction vector $\textbf{d}$, starting from parameter vector $\bm\theta$:
\begin{equation}\label{eq:scan_1D}
    f(t) = \mathcal{L}\left(\bm\theta + t\textbf{d}\right).
\end{equation}
An extension equipped with a natural interpretation is the \textit{interpolation scan} between two parameter vectors $\bm\theta, \bm\theta^\prime$ such that $\textbf{d} = \bm\theta^\prime - \bm\theta$:
\begin{equation}\label{eq:interpolation_1D}
    f(t) = \mathcal{L}\big((1-t) \bm\theta  + t \bm\theta^\prime\big).
\end{equation}
For interpolation scans, it is especially important to respect the parameter symmetries discussed in Sec.~\ref{apx:periodic_wrap} if the endpoints are not explicitly chosen to be close in parameter space. Because 1D scans can miss relevant features and curvature in the space, we recommend them to be used in addition with techniques that provide informed scan directions, such as eigenvectors of the Hessian at a given point, \ref{apx:hessian}) and the Nudged Elastic Band (\ref{apx:elastic_band}).

\subsubsection*{2D scans}
2D scans are an extension of 1D scans which can be more informative but at the cost of having a quadratic scaling in computational cost with respect to the desired grid resolution. A natural choice for 2D scans around a parameter vector $\bm\theta$ (for example a local minimum) is a linear scan with direction vectors $\textbf{d}_1, \textbf{d}_2$ and their respective interpolation parameter $t_1, t_2$:
\begin{equation}\label{eq:scan_2D}
    f(t_1,t_2) = \mathcal{L}\left( \bm\theta + t_1\textbf{d}_1 + t_2\textbf{d}_2 \right).
\end{equation}
Even though linear scans can provide valuable information about the loss landscape, they can be limited in how well they represent the relevant landscape of a VQA. ``\textit{Relevant} landscape'' refers to regions of the loss landscape which are realistically explored by VQA instance. Especially when optimizing a loss function with gradient descent, the algorithm would not interact with particularly high loss sections of the landscape. This can be partly circumvented by carefully selecting the origin points of such scans, i.e. by choosing local minima, or by combining them with the techniques presented in the following sections.

\subsection{Principal Component Analysis}\label{apx:pca}
Principal Component Analysis (PCA)~\cite{Pearson1901pca} is a tool for statistical analysis of high-dimensional data and can be used for dimensionality reduction. In contrast to modern deep learning algorithms, for example autoencoders~\cite{Schmidhuber2014overview}, the dimensionality reduction by PCA is a linear transformation. The output of the PCA algorithm are called \textit{principal components} and they form a hierarchy of orthogonal vectors. The transformation applied to a collection of data $X$ is applied via
\begin{equation}
    X^* = XW,
\end{equation}
where the linear transformation matrix $W$ is an orthogonal matrix consisting of the principal components as column vectors.
In other words, PCA applies a linear projection $W$ from a high-dimensional space onto the \textit{principal components} of the data $X$. The first principal component is the dimension in which the data $X$ shows the largest variance; higher components explaining successively less variance. Consequently, $W$ does not have to be a square matrix, i.e. not all principal components need to be calculated. However, the PCA algorithm minimizes the error $||X - X^*W^T||^2$.
PCA has varied applications in the visualization of loss landscapes and synergizes well with other techniques presented here.

\subsubsection*{Principal Components as Scan Directions}
To select among the most representative directions for a scan around a found local minimum  ($\textbf{d}$ in Eq.~\ref{eq:scan_1D} or $\textbf{d}_1, \textbf{d}_2$ in Eq.~\ref{eq:scan_2D}), one can use the first or first two principal components of the PCA algorithm applied to the optimization trajectory of parameter vectors. For various plots throughout this work, we have chosen the principal components as the scan directions. When performing 2D scans, the PCA-transformed parameters can be projected on top of the scanned loss landscape. Although one could choose random direction vectors to scan around a local minimum and project the optimization trajectory on top, the projection would likely have significantly lower variance~\cite{li2017visualizing}. The approximation error in scans such as Fig.~\ref{fig:QCBM_landscape} or Fig.~\ref{fig:NOISE_optimizer} depends on the ratio of total data variance explained in the first two PCA components. We note that this error lies only in the projection of the points onto the scanned landscape and not the 2D scan itself.\\

\subsubsection*{Dimensionality of the Optimization Trajectories}
It has been shown for highly over-parameterized classical neural networks that optimization trajectories are generally very low-dimensional. In fact, it is typical for the first two PCA components of such trajectories to explain 40\%-70\% of the total variance explained~\cite{li2017visualizing}. Considering the millions of parameters used in such models, this was unexpected but appears to help explain the effectiveness of simple stochastic gradient descent techniques on such models. For parameterized quantum circuits, to our knowledge, paralleling results have not been studied in-depth. PCA offers a direct way of evaluating the dimensionality of the optimization trajectories and thus infer information about the loss landscapes. With many of our numerical simulations, we have found that gradient descent trajectories in VQA problems can be surprisingly low-dimensional, with common variance rations explained of $>80\%$ by the first one to two principal components.

\subsection{Hessian}\label{apx:hessian}
The Hessian H of a loss function $\mathcal{L}$ is the matrix of second partial derivatives 
\begin{equation}
    (\text{H} \mathcal{L})_{ij} \equiv \frac{\partial^{2} \mathcal{L}}{\partial x_{i} \partial x_{j} }.
\end{equation}
As a symmetric matrix, the Hessian requires evaluation of $n_\theta(n_\theta-1)/2 \sim O(n_\theta^2)$ second partial derivatives, where $n_\theta$ is the number of parameters of a model. While the Hessian is expensive to calculate exactly, it has been used to study the loss landscape of deep artificial neural nets~\cite{sagun2018overparametrized}, as well as for quantum classifiers~\cite{huembeli2021characterizing, sen2021hessianlens}. Unfortunately, efficient backpropagation is not available for current quantum computers and recent proposals for calculating a single second partial derivative~\cite{Mitarai2019methodology, huembeli2021characterizing} require two applications of the \textit{Parameter Shift Rule}~\cite{Mitarai2018PSR, Schuld2018PSR}(4 loss function evaluations) for each entry of the Hessian.
\subsubsection*{1D \& 2D Scans with Hessian Eigenvectors}
The eigenvectors of the Hessian can naturally be used as direction vectors for 1D and 2D scans, as discussed in Appendix~\ref{apx:simple_scans}. They can be selected according to their respective eigenvalue to scan a particular low- or high-curvature direction. When using \textit{orqviz}, the Hessian is returned along with sorted eigenvalues and eigenvectors.
\begin{samepage}
\begin{mintedbox}{python}
from orqviz.hessians import get_Hessian
from orqviz.scans import perform_1D_scan, perform_2D_scan

hessian = get_Hessian(parameter_vector, loss_function)
# the hessian comes with sorted eigenvectors
scan1D_low = perform_1D_scan(parameter_vector, loss_function, direction=hessian.eigenvectors[0])
scan1D_high = perform_1D_scan(parameter_vector, loss_function, direction=hessian.eigenvectors[-1])

scan2D_low = perform_2D_scan(parameter_vector, loss_function, direction_x=hessian.eigenvectors[0], direction_y=hessian.eigenvectors[1])
scan2D_high = perform_2D_scan(parameter_vector, loss_function, direction_x=hessian.eigenvectors[-2], direction_y=hessian.eigenvectors[-1])
\end{mintedbox}
\end{samepage}

\subsubsection*{Distribution of Eigenvalues and Eigenvalue Ratio Scan}
The eigenvalues $\lambda_i$ of the Hessian can be used to draw valuable conclusions about the loss landscape. The most prominent example is the distribution of eigenvalues, and the magnitude of the smallest/ largest eigenvalues $\lambda_{min}, \lambda_{max}$. It is known in classical neural networks~\cite{sagun2018overparametrized} (and has been reported in some quantum neural networks\cite{huembeli2021characterizing}) that the eigenvalues of the Hessian at a local minimum cluster at zero with some strong positive outliers.\\
If all eigenvalues are non-negative $\lambda_i\geq 0$, then the Hessian matrix is \textit{positive semi-definite}. In other words, the loss landscape at that point is locally \textit{convex}. As such, the ratio of the eigenvalues $\frac{\lambda_{min}}{\lambda_{max}}$ can be viewed as a metric for non-convexity~\cite{li2017visualizing}. We note that this method is the most expensive technique presented in this work and requires well informed application. To perform such a scan, one can use \textit{orqviz} with a custom function:
\begin{samepage}
\begin{mintedbox}{python}
from orqviz.hessians import get_Hessian
from orqviz.scans import perform_2D_scan, plot_2D_scan_result
from orqviz.geometric import get_random_normal_vector, get_random_orthonormal_vector

dir_x = get_random_normal_vector(num_params)
dir_y = get_random_orthonormal_vector(dir_x)

def hessian_eigenvalue_ratio_function(trial_parameter_vector):
    hessian = get_Hessian(trial_parameter_vector, loss_function)
    # the Hessian comes with sorted eigenvalues
    return hessian.eigenvalues[0]/hessian.eigenvalues[-1]

eigenvalue_ratio_scan = perform_2D_scan(parameter_vector, hessian_eigenvalue_ratio_function, direction_x=dir_x, direction_y=dir_y)
plot_2D_scan_result(eigenvalue_ratio_scan)
\end{mintedbox}
\end{samepage}
\subsubsection*{Stochastic Approximation of the Hessian}
Even for classical neural networks with efficient backpropagation, calculation of the full Hessian is expensive. However, for VQAs with analytical gradients through the Parameter Shift rule~\cite{Mitarai2018PSR, Schuld2018PSR}, this scaling can be prohibitively limiting, even for simulations limited to a few qubits. An extension to the stochastic approximation algorithm for numerical gradients~\cite{Spall1992SPSA} has been proposed to approximate the Hessian~\cite{Spall1997secondorder} via
\begin{equation}\label{eq:H_SPSA}
    H^{\tiny spsa}_K = \frac{1}{K}\sum_{k=0}^K
    \frac{\partial^2 \mathcal{L}}{\partial \Delta_{k,1} \partial\Delta_{k,2}} \cdot \left(\frac{\Delta_{k,1}\Delta_{k,2}^T + \Delta_{k,1}^T\Delta_{k,2}}{2} \right)
\end{equation}
with 
\begin{equation}
\begin{aligned}
    \frac{\partial^2 \mathcal{L}(\bm\theta)}{\partial \Delta_{k,1} \partial\Delta_{k,2}} &= \frac12\left(
    \frac{\partial\mathcal{L}(\bm\theta + \epsilon\Delta_{k,1})}{\partial\Delta_{k,2}}
    - \frac{\partial\mathcal{L}(\bm\theta - \epsilon\Delta_{k,1})}{\partial\Delta_{k,2}}
    \right)\\
    & \begin{aligned} = \frac14 \Big(&\mathcal{L}(\bm\theta + \epsilon\Delta_{k,1} + \epsilon\Delta_{k,2})\\
    - &\mathcal{L}(\bm\theta + \epsilon\Delta_{k,1} - 
    \epsilon\Delta_{k,2})\\
    - &\mathcal{L}(\bm\theta - \epsilon\Delta_{k,1} + \epsilon\Delta_{k,2})\\
    + &\mathcal{L}(\bm\theta - \epsilon\Delta_{k,1} - \epsilon\Delta_{k,2})
    \Big).
    \end{aligned}
\end{aligned}
\end{equation}
In this case, one approximates the partial derivatives by stochastic derivatives with respect to random direction vectors $\Delta_1,\Delta_2$. The stochastically approximated Hessian $H^{\tiny spsa}$ can then be averaged over \textit{K} pairs of random vectors $\Delta$. $\epsilon\ll1$ for estimation of a numerical gradient is chosen to be small.\\
While this approximation of the Hessian has been implemented for a second-order gradient descent algorithm for VQAs~\cite{Gacon2021stochastic}, its utility in visualization is still inconclusive as the approximation requires a substantial number of averages. Another issue of the stochastic Hessian is that it is incompatible with the Parameter Shift Rule for quantum circuits, which makes it less viable for noisy quantum devices. The reason is that the Parameter Shift Rule requires treating each parameterized operation in the quantum circuit individually. \\
Nevertheless, \textit{orqviz} offers access to the stochastic approximation via
\begin{samepage}
\begin{mintedbox}{python}
from orqviz.hessians import 

hessian_SPSA_approx = get_Hessian_SPSA_approx(parameter_vector, loss_function, n_reps=n_reps, eps=eps)
\end{mintedbox}
\end{samepage}
where \textit{n\_reps} are the number of repetitions \textit{K} to average the stochastic Hessian.


\subsection{Nudged Elastic Band}\label{apx:elastic_band}
The Nudged Elastic Band (NEB)~\cite{jonsson1998nudged} algorithm is based on the mechanical model of a chain of $N+2$ pivot points $p_i, i=0,...,N+1$ that are coupled through springs with spring constant \textit{k}. In our case, the points $p$ are parameter vectors $\bm\theta$. Let $\mathcal{L}$ be a loss function which can be estimated on parameter vectors $\bm\theta$. When the end points of the chain are held constant, i.e. $i=0$ and $i=N+1$, the energy $E$ of the chain can be written as
\begin{equation}\label{eq:elastic_band}
  E(p) =  \sum_{i=1}^N\mathcal{L}(p_i) + \sum_{i=0}^{N}\frac12 k \lVert p_{i+1} - p_{i} \rVert^2.
\end{equation}
To ``relax'' the chain, the pivots are evolved under the gradient of the energy, e.g. the force $F= \nabla E$. This chain of pivots is then called the \textit{Nudged Elastic Band}. To avoid the choice of the spring constant $k$ hyperparameter, one can subtract the gradient in the tangential direction $\tau$ of the chain from the full gradient 
\begin{equation}
    F_i = - (\nabla\mathcal{L}(p_i) - (\nabla\mathcal{L}(p_i)\cdot\tau_i)\tau_i)
\end{equation}
and drop the second term in Eq.~\ref{eq:elastic_band}. The tangential vector $\tau$ should be normalized and can be calibrated via
\begin{equation}
    \tau_i =
\begin{cases}
    p_{i+1} - p_i,& \text{if } x\geq 1\\
    p_i - p_{i-1},& \text{otherwise}.
\end{cases}
\end{equation}
This calibration has been found to prevent kinks in the path and ensure stability near the saddle point~\cite{Jonsson2000improved}. In addition to this specific gradient update, it is required to re-distribute the pivots along the path that is defined by the elastic band after every gradient update, emulating the restoring forces of the springs that were removed from the simulation. 
The \textit{AutoNEB} algorithm~\cite{Kolsbjerg2016AutoNEB} wraps the NEB algorithm and inserts new pivot points where the current approximation of the true path by the piece-wise linear band is not good enough. See Appendix A of Ref.~\cite{draxler2019NEB} for a detailed description of the algorithm.

\textit{orqviz} conveniently implements a version of the NEB, as well as the AutoNEB algorithm for seamless application in VQAs. We note that the percentage error tolerance ($\alpha$ in other work) for inserting new pivots is sensitive to the sign and the magnitude of typical loss values. For example, the default values in \textit{orqviz} for the relative error tolerance (default $0.2$) and absolute error tolerance (default $0$) work well for loss functions which approach zero. For applications in VQE, as in Fig.~\ref{fig:VQE_minima_path}, where the mininum the loss function is strongly negative, these default values will not work well.
\newpage
\subsection{Parameter Concentration in Different Graph Structures}\label{apx:QAOA_more}
\begin{figure}[H]
    \centering
    \includegraphics[width=1.0\linewidth]{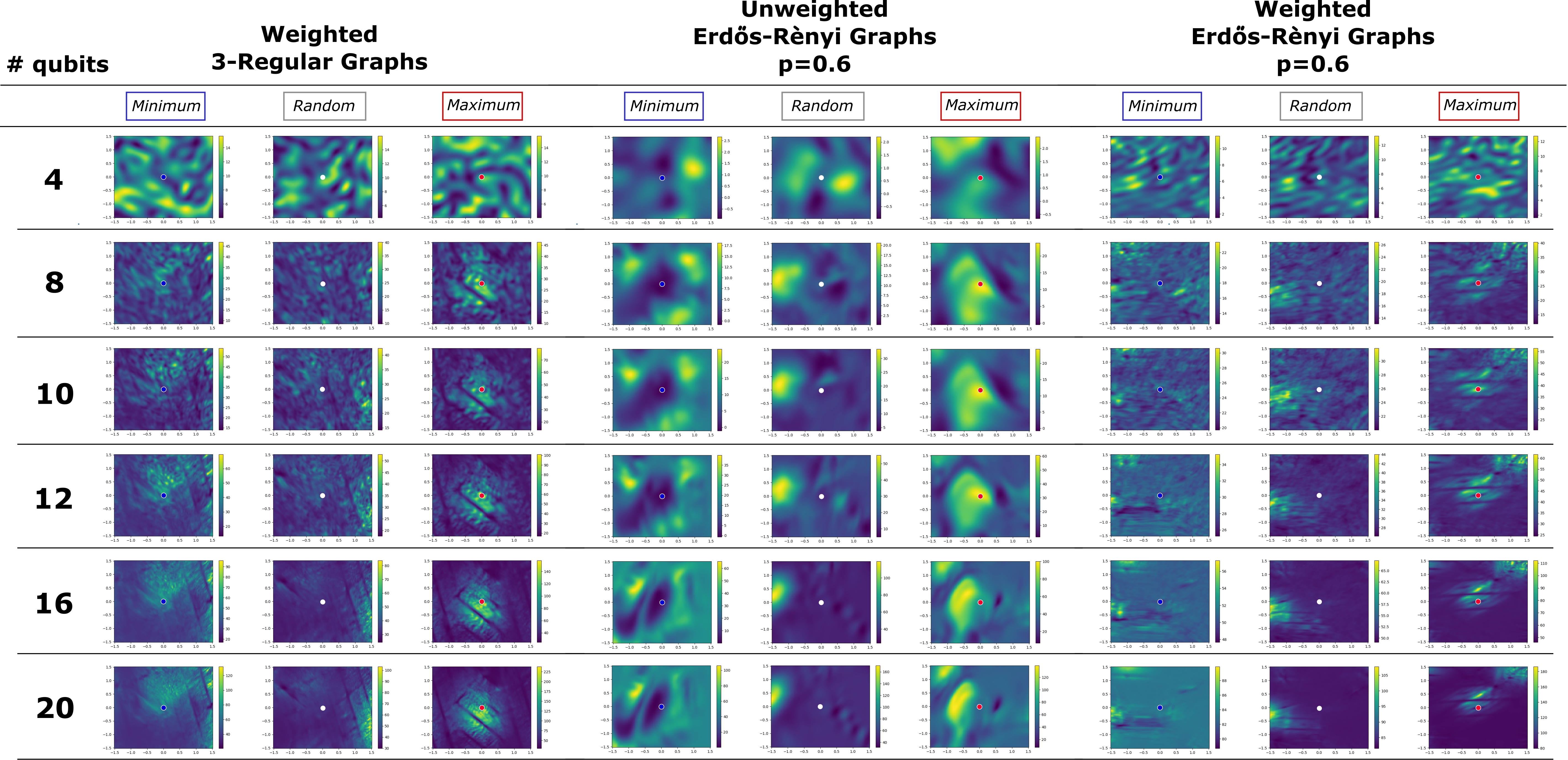}
    \caption{Parameter Concentration~\cite{brandao2018concentration} in QAOA demonstrated visually on the graph-partitioning problem (not \textsc{Max-Cut}) in analogy to Fig.~\ref{fig:QAOA_scans}. Graph-partitioning refers to splitting a graph into two sub-graphs which have approximately the same number of nodes and the sum of weights over the edges between the two sub-graphs is minimized. See details in Eqs. $7$ to $9$ in Ref.~\cite{Lucas2014Ising}.
    3-regular graphs consist of nodes which are connected to exactly three other nodes, while Erd\H{o}s-Rènyi graphs randomly connect nodes with probability $p$ (here $p=0.6$). The graphs can be weighted (weights over edges in the range $w_{ij}\in[0,1]$) or unweighted ($w_{ij}=1$).
    The QAOA ansatz uses four layers is as in~Sec.~\ref{sec:QAOA}. While transferring parameters of minima and maxima to larger graphs in the weighted case  is less clear than in the unweighted case, the phenomenon of parameter concentration is apparent in the sense that the loss landscapes have increasingly similar shapes as the graphs grow in size. For larger graphs, the loss landscape additionally become smoother/flatter away from significant features.}
    \label{fig:appendix_QAOA_more}
\end{figure}

\end{document}